\documentclass[11pt]{article}
\usepackage{jheppub}
\usepackage{amsmath,amssymb,amsfonts,graphicx,slashed,amsthm,mathtools,upgreek, enumerate, tensor, subfig}
\usepackage[dvipsnames]{xcolor}
\usepackage{arydshln}

\usepackage{comment}
\usepackage{hyperref}
\usepackage[utf8]{inputenc}
\usepackage[titletoc]{appendix}

\usepackage{cleveref}

\numberwithin{equation}{section} 
\allowdisplaybreaks

\allowdisplaybreaks

\newcommand{\be}{\begin{equation}}
\newcommand{\ee}{\end{equation}}
\newcommand{\f}{\frac}
\newcommand{\s}{\sqrt}
\newcommand{\p}{\partial}

\newcommand{\bea}{\begin{eqnarray}}
\newcommand{\eea}{\end{eqnarray}}
\newcommand{\ba}{\begin{align}}
\newcommand{\ea}{\end{align}}

\newcommand{\la}{\langle}
\newcommand{\ra}{\rangle}
\newcommand{\beq}{\begin{equation}}
\newcommand{\eeq}{\end{equation}}

\title{
Entanglement between two evaporating black holes
}

\author[a]{Akihiro Miyata}
\author[b,c]{\! Tomonori Ugajin}

\affiliation[\,a]{Institute of Physics, University of Tokyo, Komaba, \\ Meguro-ku, Tokyo 153-8902, Japan}
\affiliation[\,b]{Center for Gravitational Physics,
Yukawa Institute for Theoretical Physics, Kyoto University,
Kitashirakawa Oiwakecho, Sakyo-ku,
Kyoto 606-8502, Japan}
\affiliation[\,c]{The Hakubi Center for Advanced Research, Kyoto University,
Yoshida Ushinomiyacho, Sakyo-ku, Kyoto 606-8501, Japan}

\emailAdd{miyata@hep1.c.u-tokyo.ac.jp}
\emailAdd{tomonori.ugajin@yukawa.kyoto-u.ac.jp}

\abstract{
We study  a thermo-field double type  entangle state on   two disjoint gravitating  universes, say A and B, with an  eternal black hole on each.  As was shown previously,  its entanglement entropy of the universe A is computed by the generalized entropy on a new spacetime  constructed by suitably gluing the  black holes on A and B. We study such spacetime gluings  when universes are asymptotically flat and AdS cases, especially when the masses of these black holes  are different. We also clarify the rule to construct such a glued spacetime in  more general settings from the gravitational path integral view point.
}


\preprint{UT-Komaba/21-5}

\begin{document}

\maketitle

\parskip=10pt

\section{Introduction}

\begin{figure}[ht]
\begin{center}
\includegraphics[scale=0.3]{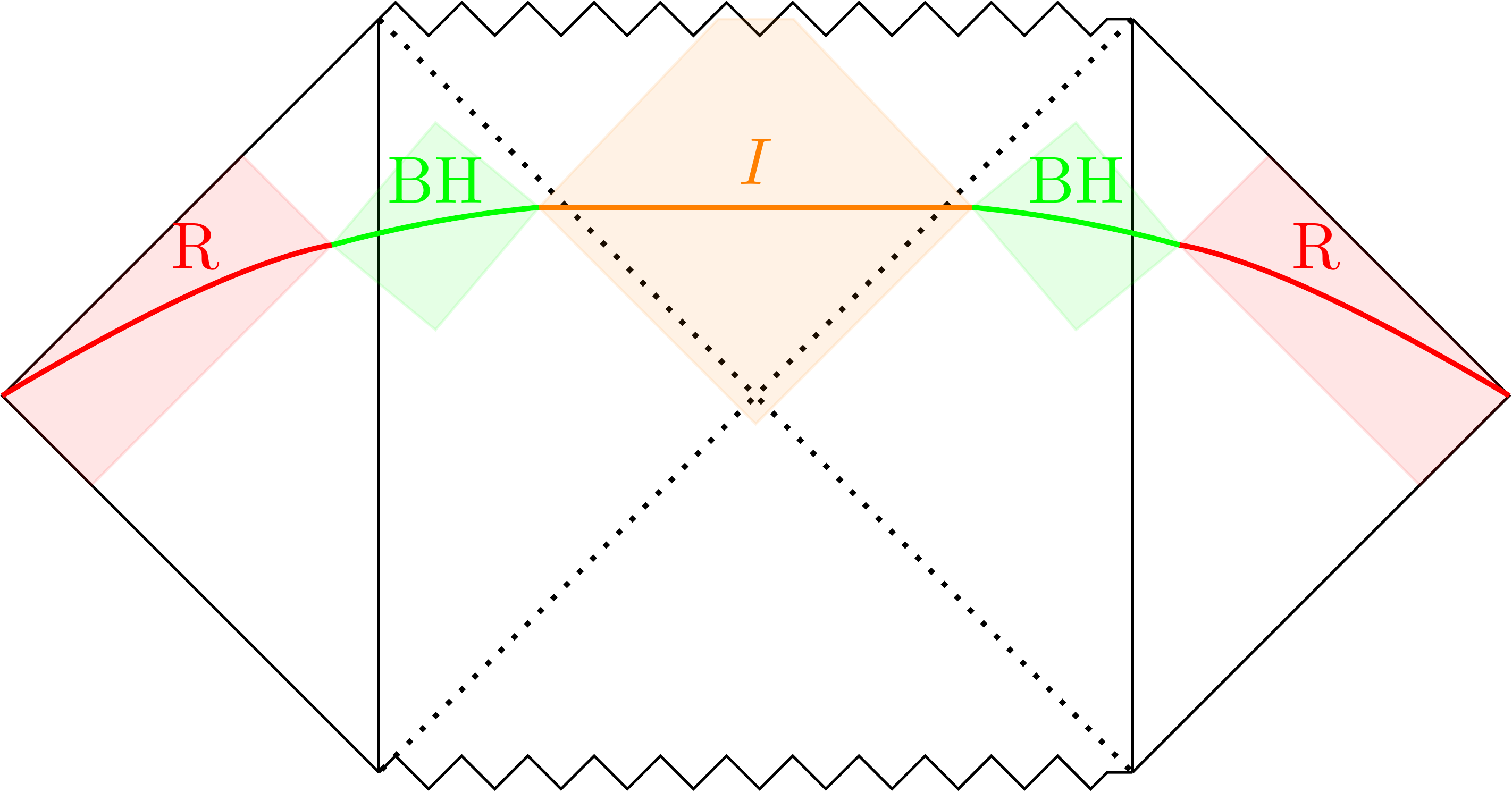}
\end{center}
\caption{The Penrose diagram of an AdS black hole coupled to the non-gravitating external bath. We take the radiation region $R$ (red solid line) in the non-gravitating external bath. After the Page time, the island region $I$ (orange solid line) becomes non-empty, and the entanglement wedge of the black hole $BH$ is outside the horizon of the black hole (green shaded region).
}
\label{fig:island}
\end{figure}

The semi-classical description of a black hole and the Hawking quanta entangled with it has appeared to be inconsistent with principles of quantum theory in many respects \cite{Hawking:1975vcx,Hawking:1976ra}. In particular, although the von Neumann entropy of the Hawking radiation should follow a Page curve \cite{Page:1993wv,Page:2013dx} due to the unitarity of quantum theory, the naive von Neumann entropy of the Hawking radiation in the semi-classical description does not. 
 Recent developments show that we can get the von Neumann entropy of the Hawking radiation $R$ in a semi-classical way by using the so-called island formula \cite{Penington:2019npb,Almheiri:2019psf,Almheiri:2019hni},
\begin{equation}
	S_{\rm{island}}(\rho_{R})= \underset{I}{\rm{MinExt}}\left[ \f{ {\rm Area}(\partial I)}{4G_{N}}+S_{\rm{eff}}(R\cup I) \right],
	\label{eq:IslandFormula}
\end{equation}
where ${\rm Area}(\partial I)$ is the area of the endpoints of a new region $I$ called the island, and $S_{\rm{eff}}(R\cup I)$ is the von Neumann entropy of the  bulk effective quantum field theory on the union of the two regions $R$ and $I$, computed on a fixed background spacetime, and $\underset{I}{\rm{MinExt}}$ denotes the extremization and the minimization of the generalized entropy over all possible islands $I$ (see figure \ref{fig:island}).
The island contribution originates from so-called replica wormholes connecting different replicas in computing the von Neumann entropy of the Hawking radiation by using the replica trick in the semi-classical description of gravity  \cite{Penington:2019kki,Almheiri:2019qdq}. 
See also
 e.g., \cite{Rozali:2019day,Chen:2019uhq,Bousso:2019ykv,Almheiri:2019psy,Chen:2019iro,Balasubramanian:2020hfs,Hollowood:2020cou,Alishahiha:2020qza,Chen:2020uac,Geng:2020qvw,Chandrasekaran:2020qtn,Li:2020ceg,Bousso:2020kmy,Dong:2020uxp,Hollowood:2020kvk,Chen:2020jvn,Chen:2020tes,Hartman:2020khs,Balasubramanian:2020coy,Balasubramanian:2020xqf,Ling:2020laa,Chen:2020hmv,Bhattacharya:2020uun,Harlow:2020bee,Akal:2020ujg,Hernandez:2020nem,Matsuo:2020ypv,Akal:2020twv,Numasawa:2020sty,KumarBasak:2020ams,Geng:2020fxl,Deng:2020ent,Karananas:2020fwx,Wang:2021woy,Kawabata:2021hac,Fallows:2021sge,Bhattacharya:2021jrn,Kim:2021gzd,Anderson:2021vof,Miyata:2021ncm,Wang:2021mqq,Ghosh:2021axl,Aalsma:2021bit,Geng:2021iyq,Balasubramanian:2021wgd,Uhlemann:2021nhu,Qi:2021sxb,Kawabata:2021vyo,Chu:2021gdb,Langhoff:2021uct,Lu:2021gmv,Akal:2021foz,Balasubramanian:2021xcm,Ahn:2021chg,Miyaji:2021lcq,Matsuo:2021mmi} for discussions on the island formula, and e.g., \cite{Chen:2020ojn,Hsin:2020mfa,Engelhardt:2020qpv,Karlsson:2020uga,Goto:2020wnk,Hirano:2021rzg,Dong:2021oad} on replica wormholes.

A frequently used setup for studying the von Neumann entropy of Hawking radiation  consists of a black hole in anti-de Sitter (AdS) spacetime and a non-gravitating external bath system  attached at  the AdS boundary (see figure \ref{fig:island} again) with the transparent boundary conditions \cite{Penington:2019npb,Almheiri:2019psf,Almheiri:2019hni,Almheiri:2019yqk,Almheiri:2019qdq}. Using such a setup, we can actually get the Page curve by using the island formula.

The introduction of the non-gravitating bath helps  to define the Hawking radiation $R$ without ambiguities coming from gravitational effects and simplifies the situation. However, if we consider a bath  which is also gravitating, then the situation drastically changes and becomes more complex. Such setups are studied in \cite{Geng:2020fxl}, see also \cite{Geng:2021iyq,Anderson:2021vof,Balasubramanian:2021wgd,Balasubramanian:2021xcm}. 
However, there are subtle points in defining the von Neumann entropy in the presence of dynamical gravity. 
One of them is that the tensor factorization of Hilbert spaces associated with gravitating regions is not well-defined 
because of the existence of the gravitational edge modes living in the boundary of such regions \cite{Donnelly:2016auv}, see also \cite{Raju:2021lwh} for a recent discussion on the implication of this fact to the island formula.
The other difficulty is that the diffeomorphism invariance of gravity does not allow us to define a region in an unambiguous way.

\begin{figure}
\begin{center}
\begin{minipage}[t]{1\hsize}
\includegraphics[scale=0.3]{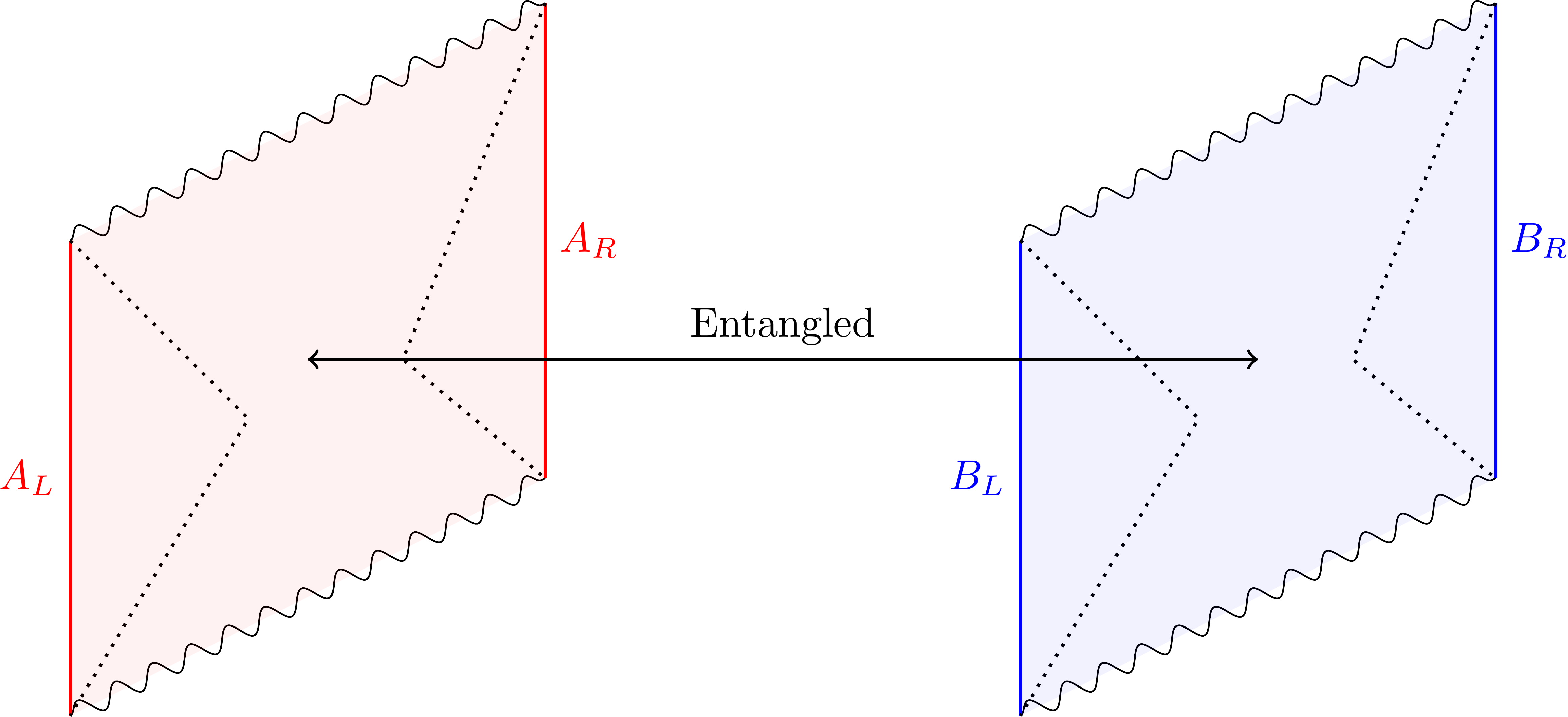}
\end{minipage}
\vspace{0.5cm}
\end{center}
\begin{minipage}[t]{1\hsize}
\includegraphics[scale=0.3]{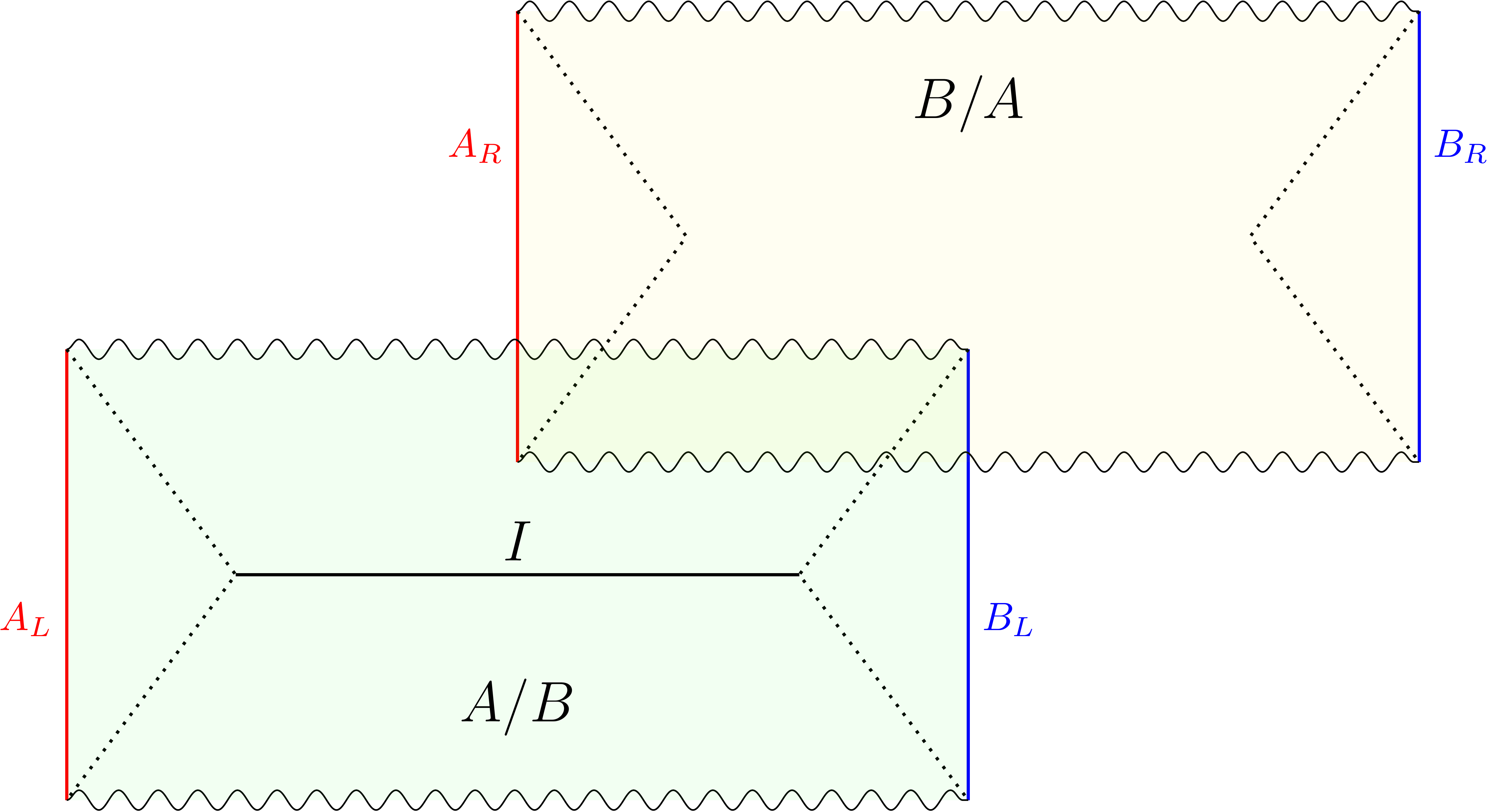}
\end{minipage}
	    \caption{Schematic picture of two phases of entangled two universes (AdS) $A,B$. (Top) The two universes are disjoint and just entangled with each other. In this case, the entanglement entropy  of the state (\ref{eq:tfdonab}) on the universe $A$ is the Hawking-like thermal entropy $S_{\rm th}(\beta)$. (Bottom) A sufficiently large entanglement between the two universes induces the geometric connection of the two spacetimes, resulting in two glued spacetimes $A/B$ and $B/A$. Unlike the above disjoint universes, since there is an island in the glued spacetime $A/B$, the entanglement entropy of the universe $A$ is given by the black hole entropy of  the glued spacetime $A/B$.}
    \label{fig:Disco}
\end{figure}

One way to avoid these subtleties is to prepare two disjoint gravitating universes $A$ and $B$, then make them entangled \cite{Balasubramanian:2021wgd}\footnote{See also  \cite{Anderson:2021vof} for the related discussion, and  \cite{Balasubramanian:2020coy,Balasubramanian:2020xqf,Miyata:2021ncm} for the discussion on entangled two disjoint universes, one of which is non-gravitating and the other is gravitating and contains a black hole.}. In this setup, one can take the entangled state to be the thermo-field double state on the two universes
\begin{equation}
	|\Psi\ra = \sum_{i}\sqrt{p_i} |\psi_i \ra_{A} \otimes |\psi_i \ra _{B}, \quad p_{i}=\f{e^{-\beta E_i}}{Z(\beta)},
	\label{eq:tfdonab}
\end{equation} 
where $ |\psi_i \ra_{A,B}$ are bulk effective field theory  eigenstates with energy $E_i$. We will study this state in detail in next section \ref{sec:EntanGravi}. In computing the entanglement entropy of the state on the universe $A$ (or the universe $B$) by using the replica trick, the dominant replica wormhole contribution is the one connecting  all the gravitating universe's replicas when the entanglement between the two universes is sufficiently large $\beta \to 0$ \cite{Balasubramanian:2021wgd}. Such a contribution leads us to the formula,
\begin{equation}
S(\rho_{A})=\rm{Min} \{S_{\rm th}(\beta),\; S_{\rm swap}(\rho_{A})\}, \quad S_{\rm swap}(\rho_{A})= \underset{I}{\rm{MinExt}}\left[ \f{ {\rm Area}(A/B,\partial I)}{4G_{N}}+S_{\rm{eff}}(I) \right],\label{eq:swap}
\end{equation}
where ${\rm Area}(A/B,\partial I)$ is the area of $\partial I$ on a new spacetime denoted by $A/B$, which is constructed by glueing two original universes $A$ and $B$, and $S_{\rm{eff}}(I)$ is the von Neumann entropy in the bulk quantum field theory of  the region $I$  on the new spacetime $A/B$ (see figure \ref{fig:Disco}). Also $S_{\rm th}(\beta) $ denotes the thermal entropy of the same bulk QFT with  the temperature $1/\beta$.  The appearance of the new spacetime $A/B$ comes from the above mentioned wormhole, and one can consider such a new spacetime $A/B$ as a realization of the ``ER=EPR" \cite{Maldacena:2013xja}, which states that the existence of the entanglement between two distant regions is related to the existence of a wormhole connecting the two regions. In other words, the sufficiently large entanglement between the two disjoint universes $A,B$ induces a geometric connection between them, leading to the introduction of the new spacetime $A/B$ (see figure \ref{fig:Disco} again).

In the previous paper \cite{Balasubramanian:2021wgd}, only the simplest setup where  the universes $A,B$ have eternal black holes with the same black hole mass (or equivalently a same black hole entropy) was studied. The resulting glued spacetime $A/B$ also contains eternal black holes with the same mass.

In this paper, we generalize this entropy calculation,  and discuss  more involved setups where the black hole mass in the universe $A$ is different from the  one  in the universe $B$,   to further understand the detailed properties of the formula \eqref{eq:swap}. In such situations, there are several  possibilities for the glued geometry $A/B$, which are indistinguishable in the previous setup \cite{Balasubramanian:2021wgd}. Our results show that the dominant glued geometry $A/B$, which contributes the entropy formula \eqref{eq:swap} is given by the one which minimizes the total  black hole entropy in $A/B$. 
For definiteness, in this paper we focus on two dimensional dilaton gravities; Jackiw-Teitelboim gravity in two-dimensional AdS spacetime and CGHS gravity in two dimensional flat spacetime.

The outline of our paper is as follows. 
In section \ref{sec:EntanGravi}, we review and explain the results of  the previous paper \cite{Balasubramanian:2021wgd}  on the entanglement entropy between two gravitating universes. In section \ref{sec:EntADS}, we construct a dilaton profile describing the glued spacetimes $A/B$ with different black hole masses for AdS JT gravity, and in the next section \ref{sec:EntFlat} we do a similar construction for CGHS gravity. 
In section \ref{sec:EntropyInter}, by using the dilaton profiles constructed in previous sections we compute the von Neumann entropy of the universe $A$ and give its interpretation. In section \ref{sec:apglue} we discuss an approximate way to construct such a glued geometry using shock waves. In section \ref{sec:GeneralSetting}, we consider more general settings where the  each of these two eternal black holes has different  the left and the right black hole masses, $M_{AL}\neq M_{AR},M_{BL}\neq M_{BR}$, for AdS JT gravity and for flat CGHS gravity, and then compute the entanglement entropy of the universe $A$. 
In section \ref{sec:conclusion}, we summarize our results and discuss their implications and future directions.
In appendix \ref{app:ADM}, we provide ADM mass formulae for AdS JT gravity and CGHS gravity.
In appendix \ref{app:ShockJT}, we explain another method to construct glued spacetimes approximately by using shock waves for AdS JT gravity, in appendix \ref{app:ShockCGHS} we explain a similar method for CGHS gravity.

\section{Entanglement between two gravitating universes}\label{sec:EntanGravi}

In this paper, we are interested in the following setup  discussed in \cite{Balasubramanian:2021wgd}. First, we prepare two disjoint gravitating universes, say A and B with the identical  cosmological constant as in figure \ref{fig:Disco}. Furthermore we assume A and B are described by a  two dimensional  dilaton gravity theory. For simplicity, we consider two dimensional  AdS  JT gravity
\be
I^{{\rm AdS}}_{{\rm grav}}=- \f{\phi_{0}}{16\pi G_{N}} \left[ \int_{D} R +\int_{\p D} 2K \right] - \f{1}{16\pi G_{N}}\int_{D} \Phi (R-\Lambda) - \f{\Phi_{b}}{16\pi G_{N}} \int_{\p D} 2K,
\ee
when the cosmological constant is  negative, and  CGHS gravity 
 \be
 I^{{\rm Flat}}_{{\rm grav}}=   - \f{1}{16\pi G_{N}}\int_{\mathcal{M}}  \left(\Phi R -\Lambda \right) -\frac{1}{16\pi G}\int_{\partial \mathcal{M}} 2\Phi K,
 \ee
when these two universes are asymptotically flat.
In these theories,  only the dynamical degree of freedom is the dilaton $\Phi$, and we can always fix the metric part. We will especially discuss situations where each universe contains a two-sided eternal black hole.
The eternal black hole in the universe A is specified by the dilaton profile $\Phi_{A}$ and  we also have $\Phi_{B}$ for the black hole in B.

As a matter degrees of freedom coupled to the gravity degrees of freedom, we take a conformal field theory. We define it both  of these universes A and B, so that the Hilbert space of the matter degrees of freedom is bipartite. On this Hilbert space, we take the following entangled CFT state, 
\be
|\Psi \ra = \sum_{i=1}^{\infty} \s{p_{i}} | \psi_{i} \ra_{A}  |\psi_{i} \ra_{B}   \quad p_{i} = \f{e^{-\beta E_{i}}}{Z(\beta)}, \label{eq:TFD stateonAB}
\ee
where $\{| \psi_{i} \ra_{A}\}_{i=1}^{\infty}$  are energy eigenstates. We can safely assume that  $CFT_{A}$ on the universe A and 
$CFT_{B}$ has the same spectrum $\{E_{i} \}_{i=1}^{\infty}$, because this conformal field theory only couples to the metric which is non-dynamical in theories of dilaton gravity.

This setup provides an ideal toy model to study the entanglement structure of evaporating black holes emitting Hawking radiation. Imagine an observer keeps collecting the Hawking quanta emitted from an evaporating black hole in some universe, say A. Then  the observer sends these quanta to another universe, say the universe B, which means that  the universe B plays a role of a bath collecting the Hawking quanta. This operation provides a concrete way to prepare the state \eqref{eq:TFD stateonAB} in question. Specifically, increasing $1/\beta$  corresponds to increase the entanglement in the radiation state \eqref{eq:TFD stateonAB}.
Thus, this identification makes it possible to study the information loss problem of an evaporating black hole through the study of the entanglement entropy between two universes in \eqref{eq:TFD stateonAB}.  

Since the universe B can be regarded as a bath collecting the Hawking quanta, it is often assumed that the universe B is non-gravitating. In this case, the entanglement entropy of the universe A is computed by the island formula \eqref{eq:IslandFormula}.  However, for an actual  evaporating black hole in our universe, the Hawking quanta are always located in the gravitating region. This motivates us to study the setup where the universe B is also gravitating. This generalization is not only for the sake of precision. Instead, in this generalization, we expect a new physical effect, called  ER=EPR  \cite{Maldacena:2013xja} comes into play. This correspondence states that  in a theory of gravity, two entangled systems must have a geometric connection through a wormhole. One of the goals of this paper is to understand how ER=EPR affects the entropy of Hawking radiation.

We are interested in  the entanglement entropy $S(\rho_{A})$ of \eqref{eq:TFD stateonAB} on the universe A, 
\be
S(\rho_{A}) = -{\rm tr}\; \rho_{A} \; \log \rho_{A}, \quad \rho_{A} =  {\rm tr}_{B}
|\Psi \ra \la \Psi |.
\ee

In \cite{Balasubramanian:2021wgd},  this entropy  $S(\rho_{A})$ was computed through the replica trick, and the result reads,
\begin{align}
S(\rho_{A}) = \rm{Min} \{S_{\rm th}(\beta),\; S_{\rm swap}(\rho_{A})\}, \quad 
S_{\rm swap}(\rho_{A})
=\underset{\overline{C}}{\text{min ext}}\left[ \f{\Phi_{A/B} (\p \overline{C})}{4G_{N}} + S_{\beta/2} (\overline{C}) -S_{{\rm vac}} (\overline{C})\right].
\label{eq:islandg}
\end{align}

This expression  is quite  similar to 
a version of the  island formula \cite{Balasubramanian:2020coy,Balasubramanian:2020xqf,Miyata:2021ncm}, which computes the same  entropy of the same state  \eqref{eq:TFD stateonAB}  defined on a gravitating universe A and a {\it non}-gravitating universe B.  However, there is one important difference  between \eqref{eq:islandg} and the island formula. Namely  the dilaton profile $\Phi_{A/B} $ appears in  \eqref{eq:islandg} is neither the dilaton profile $\Phi_{A}$  on the universe A nor, $\Phi_{B}$ on the universe B.

\begin{figure}[ht]
  \begin{minipage}[b]{0.45\linewidth}
    \centering
    \includegraphics[keepaspectratio, scale=0.3]{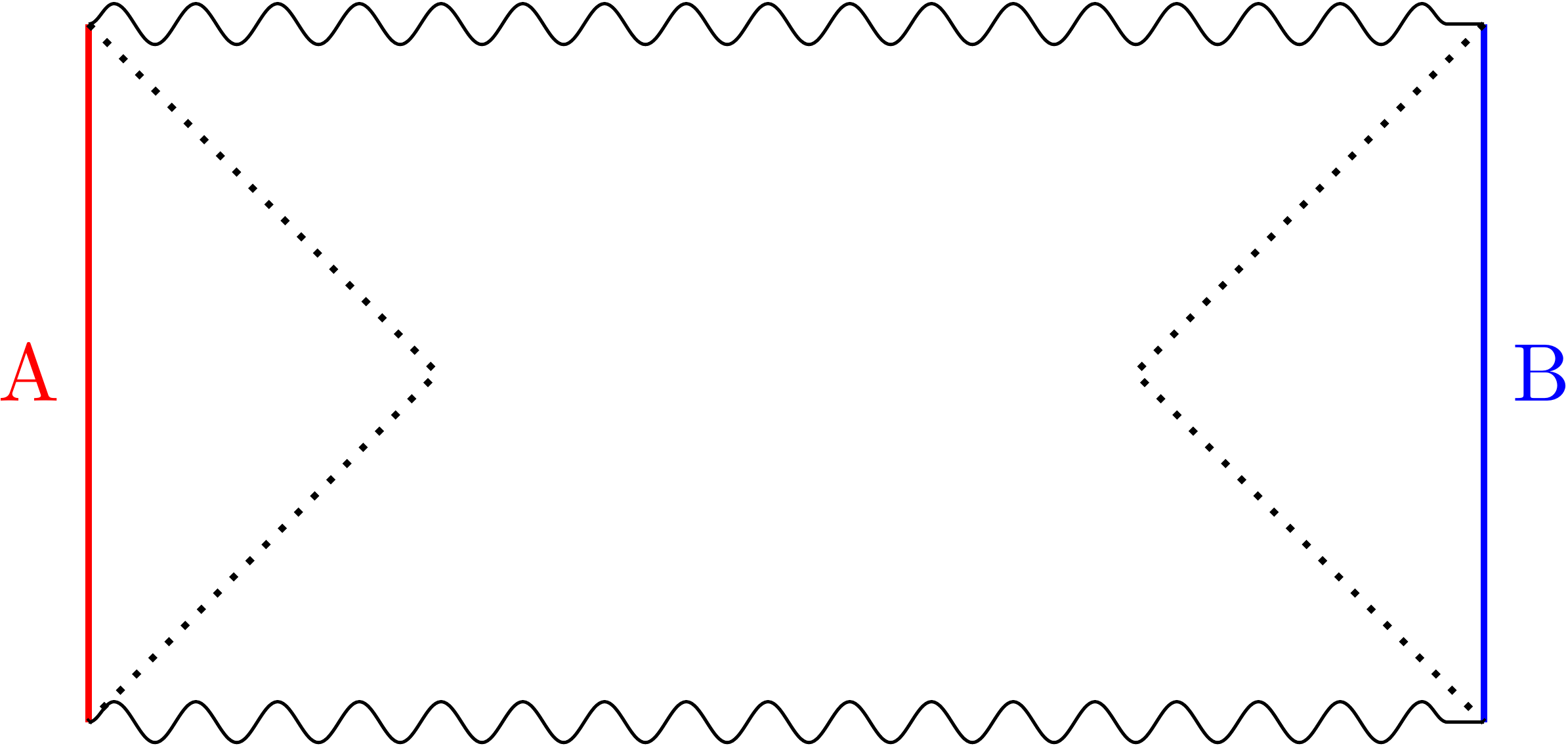}    
  \end{minipage}
  \hspace{4em}
  \begin{minipage}[b]{0.45\linewidth}
    \centering
    \includegraphics[keepaspectratio, scale=0.3]{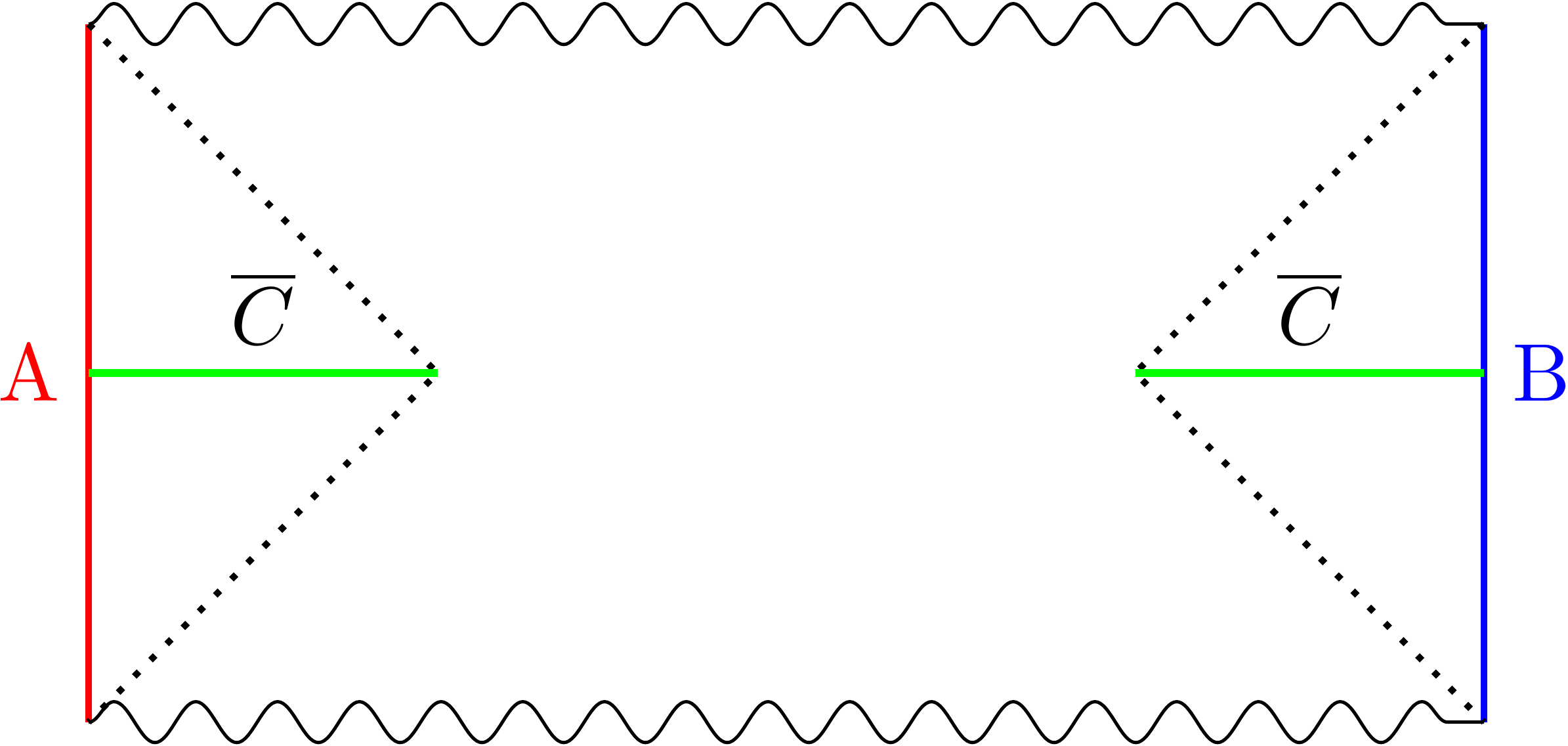}    
  \end{minipage}
  \caption{ {\bf Left}: The Penrose diagram of the new spacetime $A/B$ in AdS. The spacetime is specified by the left and right conformal boundary conditions (red and blue vertical lines).
  {\bf Right}: The same Penrose diagram of the new spacetime $A/B$ with the region $\overline{C}$ (green region) on which we evaluate the generalized entropy $S_{{\rm gen}}[\overline{C}]$.
   }
  \label{fig:PenroseGlue}
\end{figure}

Instead,  $\Phi_{A/B} $ is the dilaton profile of a {\it new} spacetime $A/B$, which is constructed by gluing two universes A and B.  For example, when A and B are both  asymptotically AdS, the spacetime geometry of each universe $\Phi_{A}, \Phi_{B}$ is specified by the boundary conditions at the left and right conformal boundaries (see figure \ref{fig:PenroseGlue}).  The dilaton profile on A/B has the boundary conditions for the universe A at its left conformal boundary, and the  boundary conditions for B at its right boundary. One of the goals of this paper is to construct this glued geometry $\Phi_{A/B}$ for a variety of examples. 

$S_{\beta/2} (\overline{C})$ in \eqref{eq:islandg} denotes the entanglement entropy of the thermal ensemble of the bulk conformal field theory, with the inverse temperature $\beta/2$, and $S_{{\rm vac}} (\overline{C})$ is the entanglement entropy of the vacuum. 

We compute the generalized entropy $S_{{\rm gen}}[\overline{C}]=\Phi_{A/B} (\p \overline{C}) + S_{\beta/2} (\overline{C}) -S_{{\rm vac}} (\overline{C})$ in the right hand side of \eqref{eq:islandg} for all possible intervals $\overline{C}$ in $A/B$ (see the right panel of figure \ref{fig:PenroseGlue}), then extremize to obtain what we call the swap entropy $S_{\rm swap}(\rho_{A})$.  The minimum between $S_{\rm swap}(\rho_{A})$ and the thermal entropy $S_{\rm th}(\beta) \rightarrow 2\pi^{2}c/3\beta, \beta \rightarrow 0$ gives the entanglement entropy $S(\rho_{A})$ for the universe A. This rule for constructing $\Phi_{A/B} $ was obtained by properly including the wormhole saddle points  of the corresponding R\'enyi entropy  ${\rm tr}\; \rho_{A}^{n}$.

\subsection*{Sketch of the derivation}

For later convenience, let us sketch the derivation of the result  \eqref{eq:islandg} for AdS black holes \cite{Balasubramanian:2021wgd}. To this end, we start from the replica trick for the entanglement entropy, 
\be
S(\rho_{A}) = \lim_{n\rightarrow 1} \f{1}{1-n}{\rm tr}\; \rho_{A}^{n}.
\ee

We then  compute the right hand side for  positive integers $n$.    From the definition of the state \eqref{eq:TFD stateonAB}, we have 
\be
{\rm tr}\; \rho_{A}^{n} =\f{1}{Z_{1}^{n}} \sum_{\{i_{k},j_{k}\}=1}^{\infty} \prod_{k=1}^{n} \s{p_{i_{k}}p_{j_{k}}} \; \la \psi_{i_{k} }|\psi_{j_{k}}  \ra_{A_{k}} \; \la \psi_{j_{k}} | \psi_{i_{k+1}} \ra_{B_{k}},
\label{eq:renyi}
\ee
where $A_{k} (B_{k})$ denotes the $k$-th copy of the universe A(B),  
with the normalization factor $Z_{1}$ defined by 
\be
Z_{1} =\sum_{i,j=1}^{\infty} \s{p_{i} p_{j}} \; \la \psi_{i} | \psi_{j} \ra_{A} \la \psi_{j} | \psi_{i} \ra_{B} .
\label{eq:normalization}
\ee

\begin{figure}[ht]
    \centering
    \includegraphics[keepaspectratio, scale=0.3]{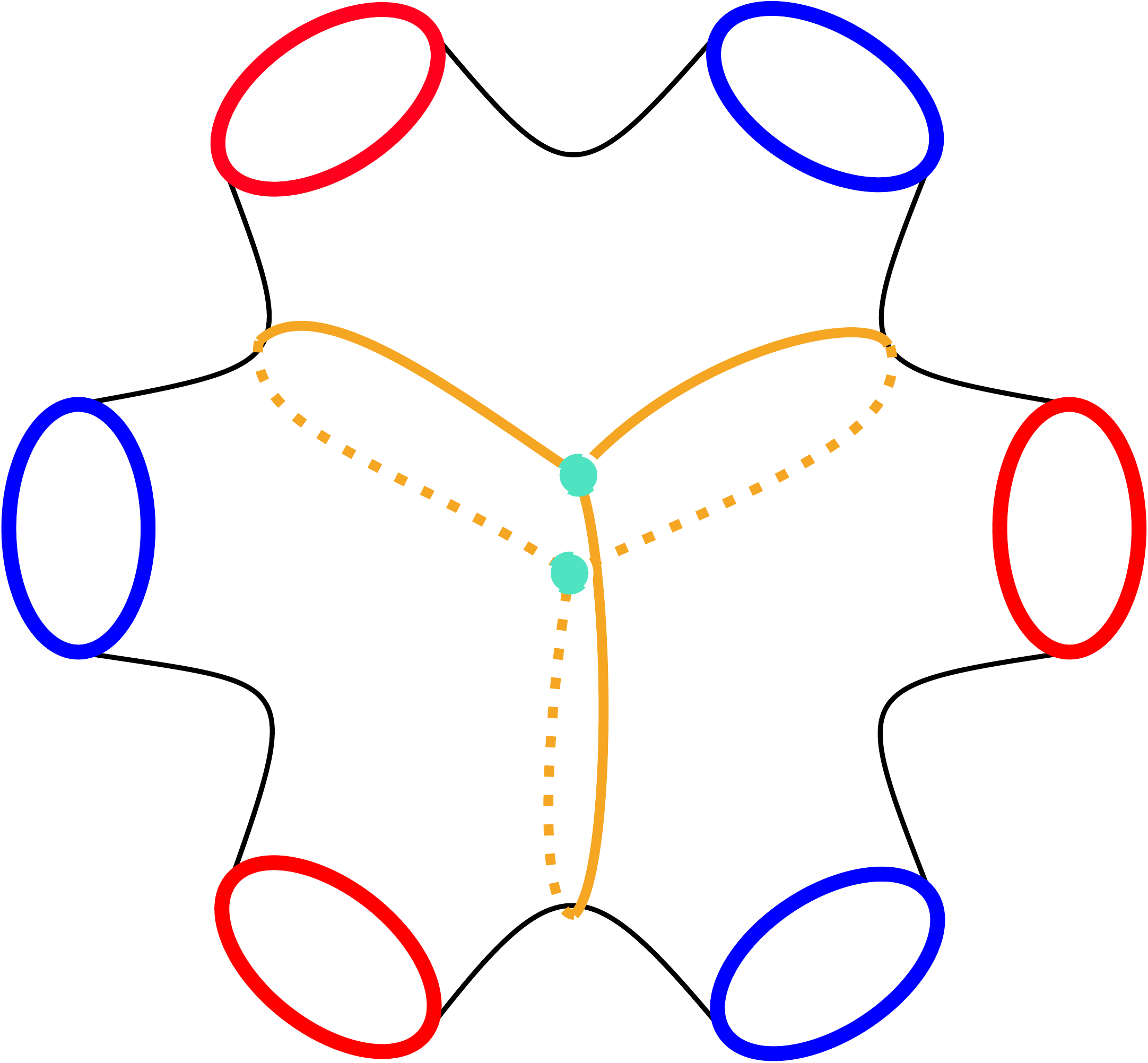}    
  \caption{Schematic picture of the replica geometry $M_{2n}$ ($n=3$ case). The $n$ ($n=3$) red boundaries  corresponds to the boundaries of the universe $A$, i.e., $\p A$. The others (blue) correspond to the boundaries of the universe $B$, i.e., $\p B$. The cyan dots are the fixed points of this replica geometry. Upon taking the quotient of the geometry by the replica symmetry $\mathbb{Z}_{N}$, the orange lines become the branch cuts.}
  \label{fig:ReplicaGeometry}
\end{figure}

The  eternal black hole state in the universe A can be prepared by a Euclidean  path integral on a half disc  with an appropriate boundary condition at the conformal boundary. We collectively denote  them by $\lambda_{A} (u)$, where $u$ is the coordinate for the boundary  circle. Similarly, we denote the boundary conditions for the black hole in B by $\lambda_{B} (u)$. 
Since each $| \psi_{i} \ra_{A_{k}}$ is a small CFT excitation on top of the fixed black hole in A, we can prepare  it  again by a path integral on the half disc with $\lambda_{A} (u)$, but now with the insertion of a CFT operator corresponding to the excitation on the south pole of the half disc. Similarly each overlap $\la \psi_{i_{k} }|\psi_{j_{k}}  \ra_{A_{k}} $  appearing in \eqref{eq:renyi} is computed  by inserting two local operators of the bulk CFT to the full disc. Altogether, the R\'enyi  entropy \eqref{eq:renyi} has a gravitational path integral expression,  with $2n$  circle boundaries $\{\p A_{k}, \p B_{k} \}$ with $4n$ CFT  local operator insertions (see figure \ref{fig:ReplicaGeometry}). 

As was shown in \cite{Balasubramanian:2021wgd}, in the high entanglement temperature limit $\beta \rightarrow 0$, the gravitational path integral is dominated by a single gravitational saddle $M_{2n}$ where all the  $2n$ boundaries $\{\p A_{k}, \p B_{k} \}$ are connected by a single wormhole. One way to think about this manifold is  first introducing a wormhole connecting  the disc $A_{k}$ to the other $B_{k}$ in the same replica to make them a cylinder (or equivalently an annulus),  then connect  these $n$ cylinders by a replica wormhole. This manifold 
therefore has a replica symmetry $\mathbb{Z}_{n}$ which shifts the $k$-th replica to the $(k+1)$-th.

\begin{figure}[ht]
    \centering
    \includegraphics[keepaspectratio, scale=0.3]{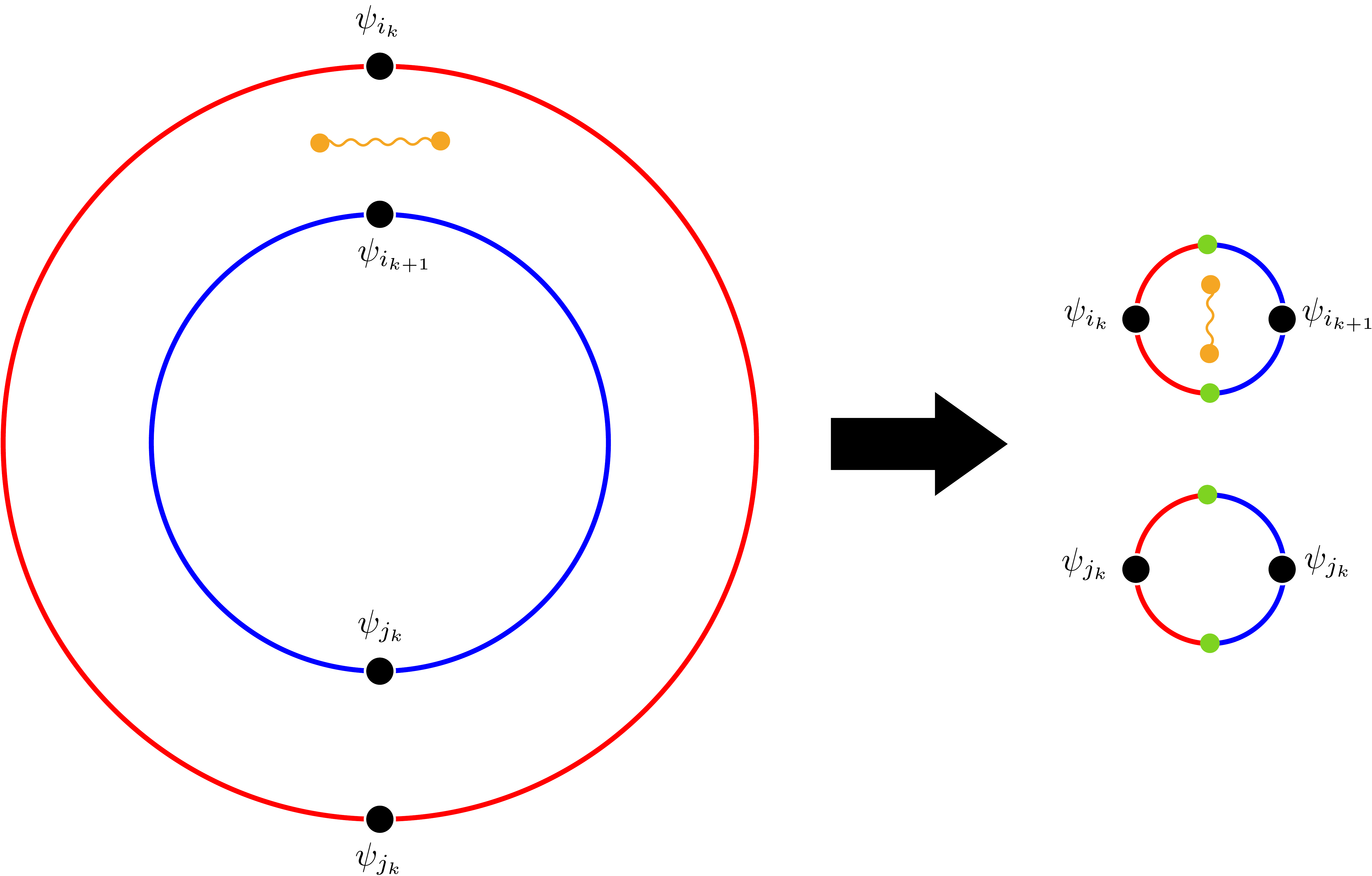}    
  \caption{ {\bf Left}: Schematic picture of the annulus geometry $M_{2n}/\mathbb{Z}_{n}$. 
  This annulus geometry has a cut (orange line) whose endpoints are the fixed points of the replica symmetry. 
  {\bf Right}: Schematic picture of the two disc geometries, which
  are obtained from the annulus geometry (left figure) by taking the $\beta \to 0$ limit. One of the two disks contains a cut (orange line). The green dots are boundary condition changing points between the two conditions of the universes $A$ and $B$.}
  \label{fig:AnnulusAndHalfDisk}
\end{figure}

The gravitational action of the saddle $M_{2n}$  is computed by the standard trick developed in \cite{Lewkowycz:2013nqa}. Since the saddle has a replica symmetry $\mathbb{Z}_{n}$, one can take its quotient  $\tilde{M}_{2n} =M_{2n}/\mathbb{Z}_{n}$, which has the topology of an annulus with a cut,  whose two boundary circles can be identified with the boundary of the universe A and B. 
Including the CFT operators, this path integral on the annulus  can be written as the thermal correlator 
\be
{\rm tr}\left[e^{-\beta_{{\rm ann}}H/2}  \psi_{i_{k}} \psi_{i_{k+1}} e^{-\beta_{{\rm ann}}H/2}  \psi_{j_{k}} \psi_{j_{k}} \right]
\ee
where $\beta_{{\rm ann}}$ denotes the circumference of the annulus, which should be distinguished from the entanglement temperature $\beta$ (see figure \ref{fig:AnnulusAndHalfDisk}). When the entanglement between A and B is strong $\beta \rightarrow 0$, it is natural to expect the renormalized length between two boundaries $\{\p A_{k}, \p B_{k} \}$  becomes shorter and shorter. This means that  if we fix its  spatial size, the  circumference gets longer, $\beta_{{\rm ann}} \rightarrow \infty$. In this limit, one can replace $e^{-\beta_{{\rm ann}}H/2} $ to the projection operator $| 0 \ra \la 0|$, thus the annulus amplitude is split into two disc path integrals, one is with a cut, and the other without. See again figure \ref{fig:AnnulusAndHalfDisk}.  The path integral on the disc  without the cut cancels with the path integral of $Z_{1}$  \eqref{eq:normalization} in the denominator of\eqref{eq:renyi}. Finally, the $n \rightarrow 1$ limit the remaining path integral on the disc with the cut   can be identified with the generalized entropy on the left hand side of \eqref{eq:islandg}. This manifests the origin of the appearance of $\Phi_{A/B}$ in \eqref{eq:islandg}.  As is clear from figure \ref{fig:AnnulusAndHalfDisk} on  the half boundary of this disk, we impose the boundary conditions of the universe A and on the other half we impose the condition for B.

\section{Entangled eternal black holes in AdS with two different masses}
\label{sec:EntADS}

In this section, we discuss the construction of the  glued geometry $\Phi_{A/B}$ which computes the entanglement entropy on two gravitating universes \eqref{eq:islandg}\footnote{See \cite{Goel:2018ubv} for a related discussion.}. We focus on the case where  the universe A  is an eternal AdS black hole with the  mass $M_{A}$,  and similarly  the universe B contains an AdS eternal black hole with  $M_{B}$. We work in the  global coordinates $(\mu, \tau)$, where the AdS metric takes the following form,
\be
ds^{2} =\f{-d\tau^{2} +d\mu^{2}}{\cos^{2} \mu}. \label{eq:jtmetric}
\ee
In the coordinates, the conformal boundaries are located at $\mu =\pm \f{\pi}{2}$. The dilaton profile of the universe A is  given by 
\be
\Phi_{A} (\mu, \tau) = b_{A}\;  \f{\cos \tau}{\cos \mu},\label{eq:AdSDilaOri}
\ee
where the parameter $b_A$ is related to the black hole mass $M_A$ through 
\be
M_A = \f{b_A^2}{16\pi G \phi_b}. \label{eq:AdSDilaMass}
\ee
We can see the above relation through the ADM mass formula \eqref{eq:ADMJTExplicit}. This dilaton profile represents an eternal black hole. 
We have a similar dilaton profile $\Phi_{B}$ for B.

The dilaton profile  $\Phi_{A/B}$   which appears in the entropy functional \eqref{eq:islandg} satisfies the following boundary conditions, 
\be
\Phi_{A/B} \rightarrow \Phi_{A} , \quad \mu \rightarrow \f{\pi}{2}, \qquad {\rm and} \qquad  \Phi_{A/B} \rightarrow \Phi_{B} ,\quad  \mu \rightarrow -\f{\pi}{2}, \label{eq:AdSGluedBdycond}
\ee
thus $\Phi_{A/B}$ is the dilaton profile of a  two sided black hole with 
different left and right masses.  The dilaton profile satisfies the following equations of motion, 
\be
\begin{aligned}
\f{1}{\cos^{2} \mu} \p_{+} \left[\cos^{2} \mu \p_{+} \Phi   \right] &= -8\pi G\; \la T_{++} \ra_{\beta} ,\\
\f{1}{\cos^{2} \mu} \p_{-} \left[\cos^{2} \mu \p_{-} \Phi   \right] &=  -8\pi G\; \la T_{--} \ra_{\beta} ,\\
2 \partial_{+} \partial_{-} \Phi-\frac{1}{\cos ^{2} \mu} \Phi&=16 \pi G\; \la T_{+-} \ra_{\beta},
\end{aligned}\label{eq:EOMJT}
\ee
with  the CFT stress energy tensor evaluated on the state \eqref{eq:TFD stateonAB}
\be
\la T_{++} \ra_{\beta} =\la T_{--} \ra_{\beta} =\f{c}{24\pi} \left( \f{2\pi}{\beta}\right)^{2}\equiv \la T \ra_{\beta}, \quad \la T_{\pm \mp} \ra_{\beta} =0 \label{eq:SEtensor},
\ee
where the chiral coordinates $x^{\pm}$ are defined by
\begin{equation}
	x^{\pm} = \mu \pm \tau.\label{eq:ChiralCoordiAdS}
\end{equation}

As we increase the entanglement temperature $1/\beta$, the eternal black hole \eqref{eq:AdSDilaOri} receives the back reaction from the CFT  stress energy tensor $\la T_{\pm \pm} \ra_{\beta}$. The result reads \cite{Balasubramanian:2020coy},
\be
\Phi_{A}(\mu, \tau) =\f{b_{A}}{2} \left[ \left(b_{0} +\f{1}{b_{0}} \right) -\f{2}{\pi} \left(b_{0} -\f{1}{b_{0}} \right) (\mu \tan \mu +1) \right],
\label{eq:AdSDil}
\ee
where $b_{0}$ is related to $\la T \ra_{\beta}$ by 
\be
\f{b_{A}}{\pi} \left(b_{0} -\f{1}{b_{0}} \right) = 16 \pi G \la T \ra_{\beta}.
\ee
The mass $M_{A}$ of the black hole remains unchanged by this back reaction.

Now let us specify the dilaton profile $\Phi_{A/B}$ which solves \eqref{eq:EOMJT} and satisfies the boundary conditions \eqref{eq:AdSGluedBdycond}. First, we note that the general solution of \eqref{eq:EOMJT} is given by 
\begin{align}
\Phi(\mu, \tau) &= \Phi_{0}(\mu, \tau) - 16\pi G \la T \ra_{\beta} \; \left( \mu \tan \mu +1 \right) \nonumber \\ 
&=\left(a\tan \mu +b \;\f{\cos \tau}{\cos \mu} \right)-16\pi G \la T \ra_{\beta} \; \left( \mu \tan \mu +1 \right), \label{eq:dilgen}
\end{align}
where  $\Phi_{0}$ denotes the ``sourceless" part of the dilaton, satisfying \eqref{eq:EOMJT} with the vanishing stress energy tensor $\la T_{\mu \nu} \ra=0$, and  this depends on two parameters $a$ and $b$ \footnote{Although there is an additional term, $\f{\sin \tau}{\cos \mu}$ for the general solution, we do not include such a term for simplicity. In the absence of this term, bifurcation surfaces are located on the $\tau=0$ time slice, as explained later.\label{foot:JTAddito}}.  

In the $\mu \rightarrow \f{\pi}{2}$ limit, the solution \eqref{eq:dilgen} approaches
\be
\Phi(\mu, \tau) =  b \;\f{\cos \tau}{\cos \mu} - \left( 8\pi^2 G \la T \ra_{\beta} -a \right)\tan \mu.
\label{eq:leftdil}
\ee

This describes a black hole with the mass 
\be
M_{R} = \f{1}{16 \pi G\phi_b} \left( b^{2} -\left( 8\pi^2 G \la T \ra_{\beta} -a \right)^{2} \right).
\ee 
This can be seen by directly using the ADM mass formula \eqref{eq:ADMJTExplicit} or applying an $SL(2,\mathbb{R})$ transformation to the  dilaton profile \eqref{eq:leftdil}, which brings the event horizon to the center of the space $(\mu, \tau) =(0,0)$. Then the  profile of the right Rindler wedge is that of the usual AdS Schwarzchild black hole. 

Similarly, by taking the $\mu \rightarrow -\f{\pi}{2}$ limit of \eqref{eq:dilgen}, we read off  the mass of the left black hole,
\be
M_{L} = \f{1}{16 \pi G\phi_b} \left( b^{2} -\left( 8\pi^2 G \la T \ra_{\beta} +a \right)^{2} \right)
\ee 

We identify the left mass $M_{L}$  to be the mass of the black hole in the universe A,  $M_{L}=M_{A}$,  and similarly $M_{R} =M_{B}$. These conditions fix the parameters $a,b$ in the dilaton profile \eqref{eq:dilgen} to be
\be
\begin{aligned}
    a &= \f{\phi_b}{2 \pi \la T \ra_{\beta}} \left(M_{B} -M_{A} \right)\\
    b &=\s{ 8 \pi G \phi_b  (M_A + M_B) +64 \pi^2 G^2 \la T \ra_{\beta}^2 + \f{ \phi_b^2}{4\pi^2\la T \ra_{\beta}^2} \left(M_{B} -M_{A} \right)^2 }\\
    &  \underset{\beta \to 0}{\approx} \s{ 8 \pi G \phi_b  (M_A + M_B) +64 \pi^2 G^2 \la T \ra_{\beta}^2 }.
\end{aligned}\label{eq:param}
\ee

 \subsection{The Penrose digram of the glued geometry in JT gravity}
\label{eq:PenroseJT}
 
 Having specified the dilaton profile $\Phi_{A/B}$ of the glued spacetime $A/B$, now let us study the causal structure of the black hole. We are particularly interested in the high entanglement temperature limit $\beta \rightarrow 0$. 
 
Let us first specify the locations of the bifurcation surfaces. 
These surfaces extremizes the profile $\p_{\tau} \Phi_{A/B} =\p_{\mu} \Phi_{A/B}=0$. It turns out that there are two such surfaces at $\mu = \mu_{R}, \; \mu_{L}$ on the time slice $\tau=0$. 
First, let us consider the right horizon, at $\mu =\mu_{R}, \tau=0$. As we will see,  this horizon approaches the right boundary  $\mu_{R} \rightarrow \f{\pi}{2}$ in the limit of our interest $\beta \rightarrow 0$, so near the right horizon we can approximate the dilaton profile as
 \eqref{eq:leftdil}, with the parameters
 \eqref{eq:param}.  Then we get
 \be 
 \sin \mu_{R} =\f{1}{b} \left( 8\pi^2 G \la T \ra_{\beta}-a\right). 
 \ee
 
 In this limit, the value of the dilaton profile at this surface is 
 \be 
 \Phi_{A/B} (\mu =\mu_{R}, \tau=0) \rightarrow b_{A}, \quad \beta \rightarrow 0,
\ee
where $b_{A}$ is defined in \eqref{eq:AdSDilaMass}. 
This is expected, since $\Phi_{A/B} \rightarrow \Phi_{A}$ near the right conformal boundary. Similarly in the high entanglement temperature limit $\beta \rightarrow 0$, the left horizon approaches the left boundary $\mu_{L} \rightarrow -\f{\pi}{2} $, and satisfies,
\be 
\sin \mu_{L} =\f{1}{b} \left( 8\pi^2 G \la T \ra_{\beta}+a\right). 
\ee
thus the dilaton value at the horizon is 
$\Phi(\mu_{L},0) =b_{B}$.

\begin{figure}[ht]
	\centering
\includegraphics[scale=0.3]{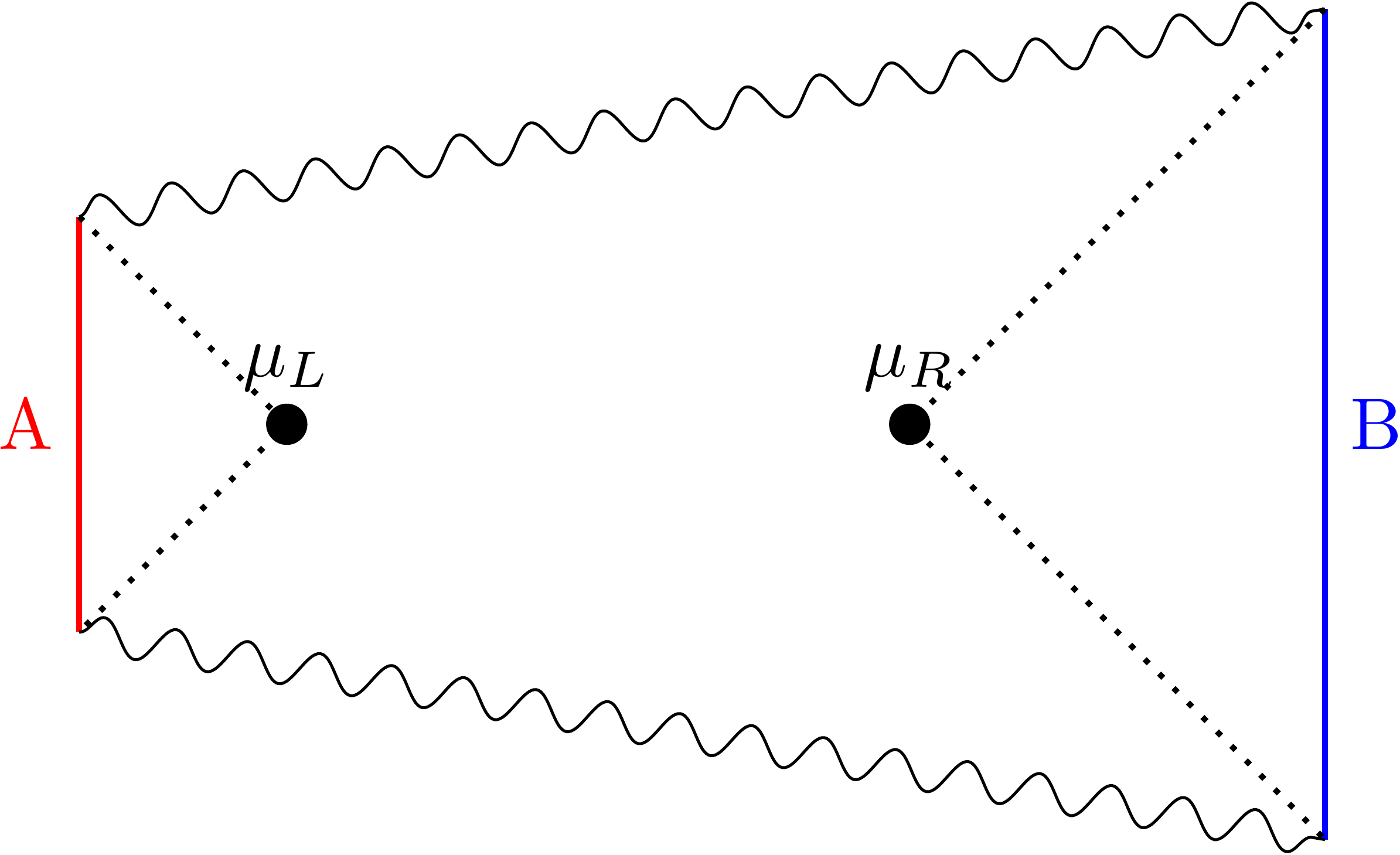}
\caption{The Penrose diagram of the glued spacetime $A/B$ with two distinct  black hole masses  $M_{A} \neq M_{B}$ in JT gravity.}
	\label{fig:penroseJTSource}
\end{figure}

Since these two horizons move toward different conformal boundaries $\mu_{R} \rightarrow \f{\pi}{2}$ and $\mu_{L} \rightarrow -\f{\pi}{2}$ as we decrease $\beta$, the black hole develops a large interior region between the two horizons. Since this  is inaccessible from  both left and right conformal boundaries, it is often  called a causal shadow region. Finally the singularity of the black hole is located at $\Phi_{A/B}(\mu, \tau) =0$. Taking into account these, we get the Penrose diagram (figure \ref{fig:penroseJTSource}) for the glued geometry $A/B$ with the dilaton profile $\Phi_{A/B}$ in JT gravity. A similar dilaton profile was discussed for example in \cite{Balasubramanian:2020coy}.

\section{Entangled eternal black holes in flat space with two different masses}\label{sec:EntFlat}

 In this section, we construct the dilaton profile $\Phi_{A/B}$ in  CGHS gravity, starting from two asymptotically flat black holes in A and B with distinct masses $M_A,\; M_B$. We begin our  discussion from  the following expression of the flat space metric,
 \begin{equation}
    \begin{aligned}
    d s^{2}&=\frac{d x^{+} d x^{-}}{\cos ^{2} x^{+} \cos ^{2} x^{-}} ,
    \end{aligned}\label{eq:metricCGHS}
\end{equation}
related to the standard form $ds^{2} = dX^{+} dX^{-}$ by the coordinate transformation $X^{\pm} = \tan x^{\pm}$. The asymptotic spatial infinity is located at $x^{+} = \pm \f{\pi}{2}$ or $x^{-} = \pm \f{\pi}{2}$. The equations of motion of CGHS gravity are given by
\begin{equation}
\begin{aligned}
-\frac{1}{\cos ^{2} x^{+} } \partial_{+}\left[\cos ^{2} x^{+}  \partial_{+}\right] \Phi &=8\pi G\left\langle T_{++}\right\rangle_\beta,\\
-\frac{1}{ \cos ^{2} x^{-}} \partial_{-}\left[ \cos ^{2} x^{-} \partial_{-}\right] \Phi &=8\pi G\left\langle T_{--}\right\rangle_\beta,\\
2\partial_{+} \partial_{-} \Phi &=16\pi G\left\langle T_{+-}\right\rangle_{\beta}-\frac{\Lambda}{2} \frac{1}{\cos ^{2} x^{+} \cos ^{2} x^{-}},
\end{aligned}\label{eq:eomForDiagoCom}
\end{equation}
 where the expectation value of the stress energy tensor $\la T_{\mu\nu} \ra_{\beta}$ is again given by \eqref{eq:SEtensor}.
 
When there is no entanglement in the state \eqref{eq:TFD stateonAB}, i.e., $\beta =\infty $, the dilaton profile for the  eternal black hole in A is 
\be
\Phi^{0}_{A} = \phi_{0,A} + \f{|\Lambda|}{4} \tan x^{+} \tan x^{-}. 
\label{eq:dila}
\ee
The mass of this black hole is given by 
\begin{equation}
	M_{A}=\frac{\sqrt{|\Lambda|}}{16 \pi G}\phi_{0, A},
\end{equation}
where we used the ADM formula \eqref{eq:ADM_GeneralDila}.
 A similar profile for the black hole in the universe B is  obtained by the replacement $\phi_{0,A} \rightarrow \phi_{0,B}$ in \eqref{eq:dila}.  This black hole has the unique bifurcation surface at $x^{\pm} =0$.
 
 As we increase the entanglement temperature $1/\beta$, these black holes receive back reaction from the CFT stress energy tensor. From  the equations of motion \eqref{eq:eomForDiagoCom}, we find the dilaton profile with the back reaction  is \cite{Miyata:2021ncm}
\begin{equation}
	\begin{aligned}
			\Phi_A(x^+,x^-)&=\phi_{0,A}+\frac{|\Lambda|}{4}\tan x^{+}\tan x^{-} - 4\pi  G \la T \ra_{\beta} \; (x^{+} \tan x^{+} + x^{-} \tan x^{-}),
	\end{aligned} \label{eq:cghsDila}
\end{equation}
where  $\la T \ra_{\beta} \equiv \la T_{\pm \pm} \ra_{\beta} $ is given by \eqref{eq:SEtensor}. Again we have a similar expression  for the universe $B$.  The mass of this black hole is given by
\begin{equation}
	M_{A} = \frac{\sqrt{|\Lambda|}}{16\pi G } \bigg( \phi_{0,A} -\frac{ \left(4\pi^2 G \la T \ra_{\beta}\right)^{2} }{|\Lambda|} \Bigg).
	\label{eq:MACGHS}
\end{equation}
$M_{A} $ decreases as we increase the entanglement temperature $1/\beta$. This is in contrast with the similar black hole in AdS JT gravity  \eqref{eq:AdSDil} whose mass kept fixed under the increase.

Having specified the dilaton profiles $\Phi_{A}$ and $\Phi_{B}$, let us construct $\Phi_{A/B}$ in CGHS model. The boundary conditions for $\Phi_{A/B}$ are given by
 \begin{equation}
	\Phi_{A/B} \to \Phi_{A} , \quad x^{\pm} \to -\frac{\pi}{2}, \qquad \text{and} \qquad \Phi_{A/B} \to \Phi_{B} , \quad x^{\pm} \to \frac{\pi}{2}.\label{eq:boundaryCondiCGHS}
\end{equation}
This  condition is analogous to the  one in JT gravity \eqref{eq:AdSGluedBdycond}. The general solution for \eqref{eq:eomForDiagoCom} is given by
 \begin{equation}
 	\begin{aligned}
 		\Phi_{A/B}(x^+,x^-)&=\Phi_0(x^+,x^-) - 4\pi  G \la T \ra_{\beta} \; (x^{+} \tan x^{+} + x^{-} \tan x^{-})\\
 		&= \left(D^{0}+\frac{|\Lambda|}{4} \tan x^{+} \tan x^{-}+D^{+} \tan x^{+}+D^{-} \tan x^{-} \right) \\
 		&	\hspace{4cm} - 4\pi  G \la T \ra_{\beta} \; (x^{+} \tan x^{+} + x^{-} \tan x^{-})
 	\end{aligned}
 	\label{eq:flatAB}
 \end{equation}
 where $\Phi_0$ is the ``sourceless" part , which satisfies the equations of motion with $\la T_{\mu\nu} \ra=0$.  The parameters $D^{0,\pm}$ in $\Phi_0$  are determined from the conditions \eqref{eq:boundaryCondiCGHS}.  We are mainly interested in $D^{+}=D^{-}\equiv D$ cases, which correspond to the situations  where  two bifurcation surfaces of the black hole are located on the $t=0$ time slice\footnote{The coordinates $x^{\pm}$ are related to $t$ by $x^{\pm}=x\pm t$.}. To determine these parameters $D^{0},D$, we focus on the asymptotic behaviors at $x^{\pm} \to \frac{\pi}{2}$ and $x^{\pm} \to - \frac{\pi}{2}$.
 In the right asymptotic  limit $x^{\pm} \to \frac{\pi}{2}$, the dilaton profile $\Phi_{A/B}$ takes the form
\begin{equation}
	\begin{aligned}
		\Phi_{A/B} \to \Phi_{R}= D^{0} + \frac{|\Lambda|}{4} \tan x^{+} \tan x^{-} + \left( D - \frac{\pi}{2} \cdot4\pi  G \la T \ra_{\beta} \right) (\tan x^{+} + \tan x^{-}) \quad \text{as } x^{\pm} \to \frac{\pi}{2},
	\end{aligned}
\end{equation}
where we introduced the notation $\Phi_R$ to distinguish it from the original one \eqref{eq:cghsDila}.

By using the ADM black hole mass formula \eqref{eq:ADM_GeneralDila}, we can see that this dilaton profile $\Phi_R$ corresponds to a black hole with the mass 
\begin{equation}
	M_{R} = \frac{\sqrt{|\Lambda|}}{16\pi G}\left( D^{0} -\frac{4}{|\Lambda|} \left( D - 2\pi^2  G \la T \ra_{\beta} \right)^{2} \right).
\end{equation}
 
 On the other hand, in the left asymptotic limit $x^{\pm} \to - \frac{\pi}{2}$, the dilaton profile $\Phi_{A/B}$ becomes
\begin{equation}
	\begin{aligned}
		\Phi_{A/B} \to \Phi_{L}= D^{0} + \frac{|\Lambda|}{4} \tan x^{+} \tan x^{-} + \left( D + \frac{\pi}{2} \cdot4\pi  G \la T \ra_{\beta} \right) (\tan x^{+} + \tan x^{-}) \quad \text{as } x^{\pm} \to -\frac{\pi}{2},
	\end{aligned}
\end{equation}
  where we again introduced the notation $\Phi_L$, and from the ADM mass formula \eqref{eq:ADM_GeneralDila} this dilaton profile $\Phi_{L}$ gives the black hole mass
 \begin{equation}
	M_{L} = \frac{\sqrt{|\Lambda|}}{16\pi G}\left( D^{0} -\frac{4}{|\Lambda|} \left( D + 2\pi^2  G \la T \ra_{\beta} \right)^{2} \right).
\end{equation}

 We identify these black hole masses $M_L$, $M_R$  with $M_A$ defined in \eqref{eq:MACGHS} and $M_{B}$ obtained by the replacement $\phi_{0,A} \rightarrow \phi_{0,B}$ in $M_{A}$ respectively, i.e,  $M_L=M_A$ and $M_R=M_B$. These identifications determine the parameters $D^{0},D$ in terms of $\phi_{0,A}$, $\phi_{0,B}$ and $\la T \ra_{\beta}$ as follows
\begin{equation}
	\begin{aligned}
		D^{0} & = \frac{\phi_{0,A} + \phi_{0,B}}{2} + |\Lambda| \frac{(\phi_{0,A} - \phi_{0,B})^2}{256 \pi^4 G^2 \la T \ra_{\beta}^2} \\
		&\underset{\beta\to 0}{\approx} \frac{\phi_{0,A} + \phi_{0,B}}{2},\\
		D & = |\Lambda| \frac{\phi_{0,B} - \phi_{0,A}}{32 \pi^2 G \la T \ra_{\beta}}.
	\end{aligned}
	\label{eq:parameters}
\end{equation}

\subsection{The Penrose diagram in the CGHS gravity}

Let us study the causal structure of  the dilaton profile $\Phi_{A/B}$ obtained by plugging \eqref{eq:parameters} into \eqref{eq:flatAB} in CGHS model. We are again interested in the high entanglement temperature limit $\beta \to 0$.

At first, we determine the locations of the bifurcation surfaces  of the black hole, at which the dilaton profile $\Phi_{A/B}$ takes extremal values. From the extremality conditions $\partial_{\pm} \Phi_{A/B}=0$, we get two distinct extremal surfaces $x_{R}^{+}=x_{R}^{-}\equiv x_{R} $, $x_{L}^{+}=x_{L}^{-}\equiv x_{L}$, which are located on the $t=0$ time slice. 

In the limit $\beta \to 0$, the extremality conditions take the following simple form
\begin{equation}
	\tan x_{R} = - \frac{4}{|\Lambda|} \left( D - 2\pi^2  G \la T \ra_{\beta} \right),\quad 	\tan x_{L} = - \frac{4}{|\Lambda|} \left( D + 2\pi^2  G \la T \ra_{\beta} \right).
\end{equation}
These equations imply that, in the limit $\beta \to 0$, the right bifurcation surface approaches the right spatial infinity $x_{R}^{\pm} \to \frac{\pi}{2}$ and the left one does the left spatial infinity $x_{L}^{\pm} \to -\frac{\pi}{2}$.

\begin{figure}[ht]
     \centering
     \includegraphics[scale=0.3]{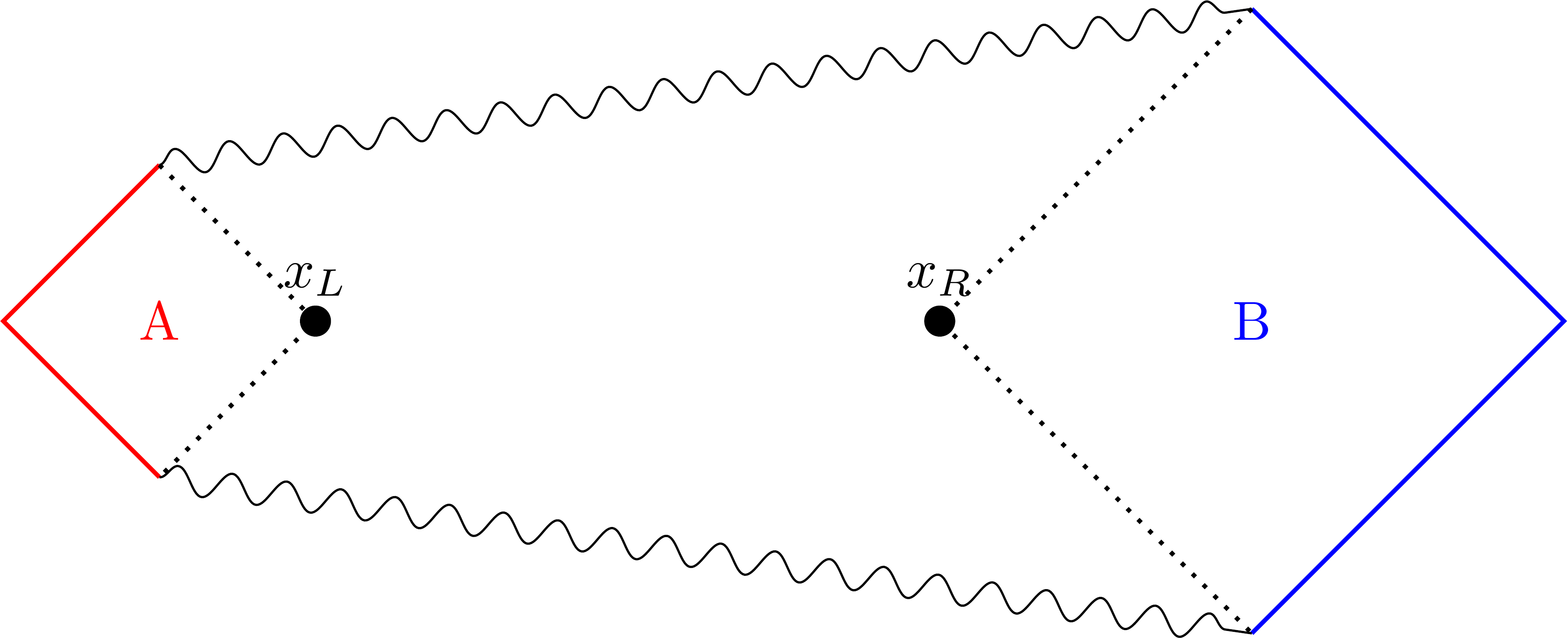}
     \caption{The Penrose diagram of the glued spacetime $A/B$ with two different black hole masses in CGHS gravity.}
     \label{fig:PenroseCGHSSource}
 \end{figure}

In this limit, the dilaton values at the bifurcation surfaces are given by
\begin{equation}
	\begin{aligned}
		\Phi_{A/B}(x^{\pm}=x_{R}) \to &  D^{0} -\frac{4}{|\Lambda|} \left( D - 2\pi^2  G \la T \ra_{\beta} \right)^{2} = \phi_{0,B} -\frac{ \left(4\pi^2 G \la T \ra_{\beta}\right)^{2} }{|\Lambda|}  ,\\
		\Phi_{A/B}(x^{\pm}=x_{L}) \to &  D^{0} -\frac{4}{|\Lambda|} \left( D + 2\pi^2  G \la T \ra_{\beta} \right)^{2}= \phi_{0,A} -\frac{ \left(4\pi^2 G \la T \ra_{\beta}\right)^{2} }{|\Lambda|} .
	\end{aligned}
	\qquad \text{as } \beta \to 0.
\end{equation}
These dilaton values at the horizons are also consistent with the discussion in the previous section. From the above discussion plus the location of the singularity at $\Phi_{A/B} (x^{+},x^{-})=0$, we get the Penrose diagram (figure \ref{fig:PenroseCGHSSource}) corresponding to the dilaton profile $\Phi_{A/B}$ in the CGHS gravity.

 \section{Entropy calculation and its interpretation}\label{sec:EntropyInter}
 
 In this section, we compute the entanglement entropy $S(\rho_{A})$ using  \eqref{eq:islandg}, by plugging the dilaton profile $\Phi_{A/B}$ of the glued geometry into the formula.  
 We take the following ansatz for $\bar{C}$ on the reflection symmetric slice, 
 \be
 \bar{C}: \left[-\f{\pi}{2}  -\f{\pi y }{2} \right]\;  \cup\;  \left[ \f{\pi x }{2} , \;  \f{\pi}{2} \right], \quad 0< x, y< 1,
 \ee
 see figure \ref{fig:PenroseGlue} for the AdS case.
 
Let us  assume the bulk CFT  has a holographic dual.  Then  the bulk entanglement entropies have the following simple expressions
\be
S_{\beta/2} [\bar{C}] = \f{c}{3} \log \left[\f{\beta}{2\pi} \sinh \f{\pi^{2}}{\beta} \left(1-\f{x+y}{2} \right)\right], \quad S_{{\rm vac}}
[\bar{C}] =\f{c}{3} \log \left[ 2 \sin \pi \left(\f{x+y}{2} \right) \right]. 
\ee
 
 The actual dilaton profile is given by \eqref{eq:dilgen} with  \eqref{eq:param} for AdS JT gravity, and  \eqref{eq:flatAB}   with \eqref{eq:parameters} for flat CGHS model. Then $S_{{\rm swap}} (\rho_{A}) $ in \eqref{eq:islandg} is computed by taking the extrema of the following function with respect to two variables $x$ and $y,$
 \be
S_{\rm gen} (x,y) = \f{\Phi_{A/B}(x)}{4G_{N}} + \f{\Phi_{A/B}(y)}{4G_{N}} +\f{c}{3} \log \left[\f{\beta}{2\pi} \sinh \f{\pi^{2}}{\beta} \left(1-\f{x+y}{2} \right)\right]-\f{c}{3} \log \left[ 2 \sin \pi \left(\f{x+y}{2} \right) \right].
 \ee
 
 In both cases, in the $\beta \rightarrow 0$ limit, the bifurcation surfaces approach the asymptotic  boundaries (or spatial infinities).
 The  value of the entropy is half of the sum of the entropies of the original black holes,
 \be
 S_{{\rm swap}} (\rho_{A}) = S_{BH_{A}} +S_{BH_{B}}
 \ee

 \begin{figure}[ht]
 \hspace{-2em}
  \begin{minipage}[b]{0.45\linewidth}
    \centering
    \includegraphics[keepaspectratio, scale=0.6]{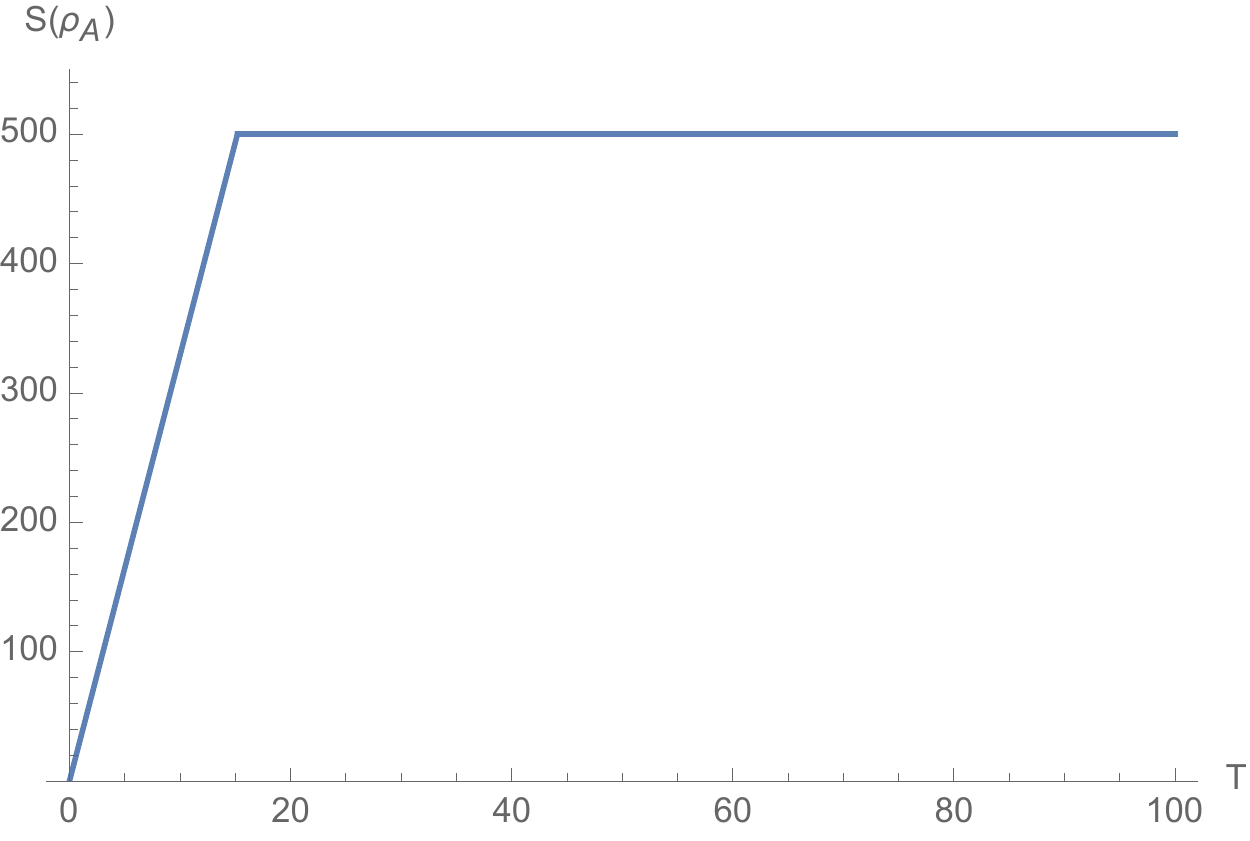}    
  \end{minipage}
  \hspace{4em}
  \begin{minipage}[b]{0.45\linewidth}
    \centering
    \includegraphics[keepaspectratio, scale=0.6]{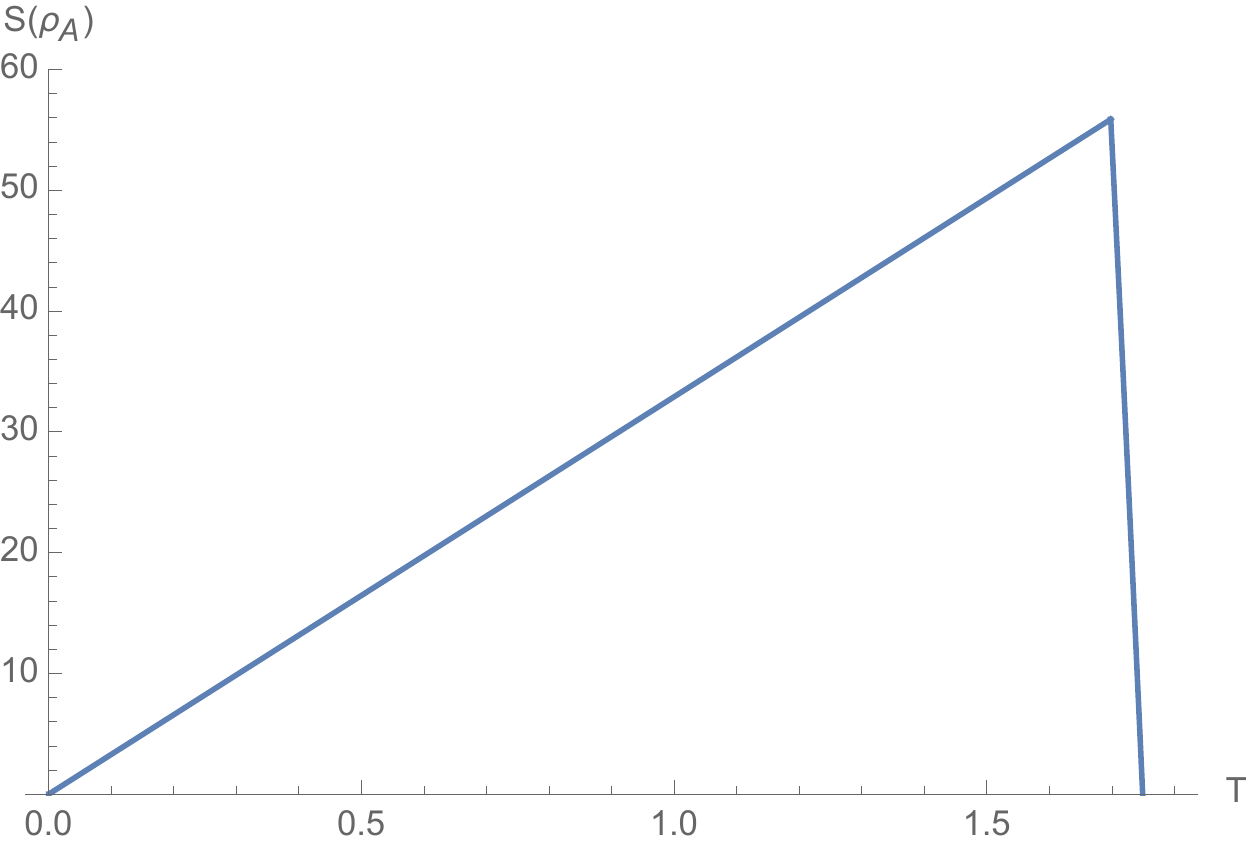}    
  \end{minipage}
  \caption{
  {\bf Left}: Plots of the entanglement entropy (\ref{eq:islandg}) of the AdS JT gravity case as a function of the entanglement temperature $T=1/\beta$. We set the parameters to be $\phi_{0,A}=300,\; \phi_{0,B}=200,\; c=10, \; 4G_{N}=1$.
  {\bf Right}: Similar Plots for flat CGHS model case as a function of $T=1/\beta$. We set the parameters to be $\phi_{0,A}=300,\; \phi_{0,B}=200,\; |\Lambda|=100,\; c=10, \; 4G_{N}=1$.
   }
  \label{fig:PagePlot}
\end{figure}
 
 We plot the resulting entanglement entropy \eqref{eq:islandg} as a function of the entanglement temperature $1/\beta$ in figure \ref{fig:PagePlot} for black holes in AdS JT gravity as well as in flat CGHS model. In both cases,  in the low temperature limit the results coincide with the thermal entropy of the bulk CFT, which reproduce the Hawking's result.  In the high temperature regime where the results are proportional to the area of the black holes, two cases show distinct behaviors. Namely, whereas for the black holes in AdS JT gravity, the entanglement entropy saturates to a constant value, for the black holes in flat  CGHS model, the entropy is decreasing.  This difference can be understood from the fact that the black holes in flat space can evaporate by emitting Hawking quanta and lose their masses, on the contrary to the black holes in AdS.

 We emphasize that our result is not  obtained by applying the island formula for the universe A and B  independently, then summing up their outcomes.  Such a contribution would come from the replica wormhole connecting $n$ copies of A and the other replica wormhole connecting Bs, without any further connection between A and B. However,  as was argued in  \cite{Balasubramanian:2021wgd} this does not yield the expected result, i.e., the sum of two island formulae for A and B. This is because , in this saddle, the bulk CFT  part of the path integral can not be interpreted as a R\'enyi entropy, due to the alignment of the operators the numerator of \eqref{eq:renyi}. Also, if we choose this saddle, then one can not kill the unbounded growth of the denominator. Thus the resulting entropy increases without any bound as well,  in the high temperature limit. 
 
  \section{Approximate gluings}
 \label{sec:apglue}
 
 We have discussed  exact solutions for  $\Phi_{A/B}$  for  the asymptotically AdS black holes \eqref{eq:AdSDilaOri} and those in flat space  \eqref{eq:cghsDila}. This was  possible  because the original dilaton profiles in A and B are simple enough to find such  interpolating solutions $\Phi_{A/B}$. Then the question arises, is there any way to approximately construct such an interpolating solution, especially when it is difficult to find an exact solution. 
In this section we propose an idea in this direction, using  shock waves in these geometries. Again, let us take the AdS black hole with the mass $M_{A}$ \eqref{eq:AdSDilaOri} and the one  with $M_{B}$ for example, and think gluing them to obtain $\Phi_{A/B}$ by an approximate mean. We start from the dilaton profile \eqref{eq:AdSDilaOri}, and then imagine inserting a left moving shock wave in the black hole interior. The expectation value of the  CFT  stress energy tensor is given by 
\be
\la T_{\pm \pm} \ra = \la T_{\pm \pm} \ra_{\beta} +\la T_{\pm \pm} \ra_{S},\quad \la T_{\pm \mp} \ra = 0,\label{eq:StresEnergyTensorWiS}
\ee
with 
\be
\la T_{+ +} \ra_{S} =E  \delta (x^{+}-x^{+}_{0}),\quad \la T_{- -} \ra_{S} =0.
\ee
 Here we also added the thermal expectation value $\la T_{\pm \pm} \ra_{\beta}$ coming from the entanglement temperature.  Now the set of equations of motion for the dilaton  is given by  \eqref{eq:EOMJT} with the above stress tensor \eqref{eq:StresEnergyTensorWiS}. The new geometry depends on two parameters, the energy $E$ of the shock wave, and its location $x^{+} =x^{+}_{0}$. As we show in appendix \ref{app:ShockJT}, the mass of the black hole to the right of the shock  is  changed to  $M_{AR} = M_{AR} (x^{+}_{0},E)$ from its original mass $M_{A}$. On the other hand, the mass of the left black hole remains to be the same, $M_{AL} =M_{A}$. The explicit form of the dilaton profile reads,
 \begin{align}
\Phi &= b_{A} \left(\f{\cos \tau}{\cos \mu} \right) -16\pi G \la T \ra_{\beta} ( \mu \tan \mu +1) \nonumber \\[+10pt]
&\qquad - 16 \pi G E  
\cos^{2} \left(\f{x^{+}_{0} + x^{-}}{2}\right) \left[ \tan \mu -\tan \left(\f{x^{+}_{0} + x^{-}}{2} \right) \right] \Theta (x^{+} -x^{+}_{0} ).
\label{eq:fulldilwiS}
\end{align}
 
 To approximate the glued geometry $\Phi_{A/B}$ by the dilaton profile with the shock \eqref{eq:fulldilwiS}, we impose the condition that the right mass  coincides with the mass of the black hole in B, $M_{AR} (x^{+}_{0},E) =M_{B}$. Let us denote the dilaton profile \eqref{eq:fulldilwiS} with this condition by $\tilde{\Phi}_{A/B}$. Details of the expression of $M_{AR} (x^{+}_{0},E)$ can  be found in Appendix \ref{app:ShockJT}. Of course, this is not enough to entirely  fix these two parameters $x^{+}_{0}, E$. However as we can immediately see, if we plug the approximate dilaton profile $\tilde{\Phi}_{A/B}$ into the generalized entropy \eqref{eq:islandg} then extremize, then it gives the correct swap entropy $S_{{\rm swap}} (\rho_{A})$ in the $\beta \rightarrow 0$ limit,  as long as the constraint is satisfied,  because the resulting entropy is  again given by the sum of two entropies of the black holes in A and B. We can also see, the only the difference between the true  dilaton profile $\Phi_{A/B}$  and the approximate one   $\tilde{\Phi}_{A/B}$  the   presence of the  term proportional to $\sin \tau/ \cos \mu$, which only shifts the location of the horizon in the timelike direction. 
 
 We have a similar construction for CGHS model, and explain it in appendix \ref{app:ShockCGHS}.

 \section{More general settings}\label{sec:GeneralSetting}
 
 So far, we have been discussing the cases where two horizons of the eternal black hole of each universe have the same masses $M_{L} =M_{R}$.  When they are different, we should carefully construct the glued spacetime $A/B$ on which we compute the generalized entropy. 

As an example of such eternal black holes with  $M_{L} \neq M_{R}$,  let us again consider the dilaton profile of the form \eqref{eq:dilgen}, which we reproduce here,
\be
\Phi_{A} (\mu, \tau)=\left(a_{A} \tan \mu + b_{A}\f{\cos \tau}{\cos \mu} \right) - 16 \pi G X (\mu \tan \mu +1).
\label{eq:dilneq}
\ee

  In previous sections, we regard this as the dilaton profile of the glued geometry $A/B$, which results from gluing two eternal black holes with the $M_{AL} =M_{AR}$, and $M_{BL} =M_{BR}$, but now we regard it as an example of the dilaton profile of the {\it }single universe, say the universe A, with  $M_{AL} \neq M_{AR}$.  Also in this setup, we are regarding  the parameter $X$  as merely a parameter, not related to the entanglement temperature $1/\beta$ and the corresponding CFT stress tensor  $\la T \ra_{\beta}$. 
  
  The masses of the left and right black holes $M_{AL}$,  $M_{AR}$ are given by 
  \be
  M_{AL} = \f{1}{16 \pi G\phi_b} \left( b_{A}^{2} -\left( 8\pi^2 G X +a_{A} \right)^{2} \right), \quad   M_{AR} = \f{1}{16 \pi G\phi_b} \left( b_{A}^{2} -\left( 8\pi^2 G X -a_{A} \right)^{2} \right).
  \ee
 
 We also have a similar dilaton profile  for the universe B by the replacement  $(a_{A}, b_{A}) \rightarrow  (a_{B}, b_{B})$ in \eqref{eq:dilneq}, while keeping $X$ intact. 
 
 Now we would like to  specify the dilaton profile of the glued geometry $\Phi_{A/B}$ which appears in the formula for the entanglement entropy 
 \eqref{eq:islandg}. In this  generalized case,  there are four candidates of the dilaton profile, namely $\Phi_{AL/BR}, \; \Phi_{AL/BL}, \; \Phi_{AR/BL},\Phi_{AR/BR}$. Here for example  we denote by  $\Phi_{AL/BR}$  the dilaton profile which approaches $\Phi_{A}$ near the left boundary and  $\Phi_{B}$ near the right boundary, satisfies the following boundary conditions,
\be
\Phi_{A/B} \rightarrow \Phi_{A} , \quad \mu \rightarrow -\f{\pi}{2} , \qquad {\rm and} \qquad  \Phi_{A/B} \rightarrow \Phi_{B} ,\quad  \mu \rightarrow +\f{\pi}{2}. 
\ee
 This dilaton profile $\Phi_{AL/BR}$  contains the left horizon of the black hole in A  and the right horizon of the black hole in B. Other dilaton profiles are also defined in a similar way, according to the choice of two horizons in A/B  out of  candidate horizons $\{AL,AR\}$ in A and $\{BL,BR\}$ in B.
 They all satisfy the equations of motion \eqref{eq:EOMJT} with $\la T_{++} \ra = \la T_{--} \ra = \la T \ra_{\beta} +X$. The difference between them  is coming from the choice of the boundary conditions near the conformal boundaries $ \mu \rightarrow \pm \f{\pi}{2}$.

One may be puzzled by the fact that we are dealing with dilaton profiles obeying distinct boundary conditions, for the reason that in a calculation of a semi-classical  gravitational path integral we always fix the boundary conditions. To clarify this point, let us come back to the replica derivation of the formula \eqref{eq:islandg} which was briefly reviewed in section \ref{sec:EntanGravi}, as in  figure \ref{fig:ReplicaGeometry}.  The entanglement entropy is computed by evaluating the on shell action of the Euclidean wormhole connecting two discs for the universes A and B, in the presence of a cut $\overline{C}$. In this description we have a unique boundary condition on $\p A$ and $\p B$.  When the masses of the left and right horizons  are not necessarily the same $M_{AL} \neq M_{AR}$,$M_{AL} \neq M_{AR}$, there are four possible types of Euclidean wormholes. Such a wormhole is constructed first by flipping the left and right of these discs, then connecting these two. For example, suppose that we flip the  disc of the universe B and then connect A and B.  Its Lorenzian continuation of the Euclidean cylinder gives two eternal black holes, one of which contains the cut $\overline{C}$. Furthermore, as is explained in \cite{Balasubramanian:2021wgd} these two eternal black holes gets  disjoint in the high entanglement temperature limit $\beta \rightarrow 0$, as depicted in figure \ref{fig:AnnulusAndHalfDisk}. This is  because if we regard the cylinder as an annulus, its circumference gets large in this limit. Therefore, for the eternal black hole with the cut, when analytically continued, the dilaton is given by $\Phi_{AR, BR}$\footnote{Notice that without any flip, we get the dilaton  $\Phi_{AR, BL}$,  because as in \cite{Balasubramanian:2021wgd} we rotate the disc B by $\pi$ relative to the disc A, to adjust the locations of the CFT operators.}. Similarly, if we flip both two universes we get a wormhole whose Lorenzian dilaton profile is given by  $\Phi_{AL, BR}$.

For each wormhole with the cut $\overline{C}$, the gravitational action is proportional to its dilaton profile $ -(n-1) \Phi [\overline{C}]\;$ in the $n \rightarrow 1$ limit. Upon taking the extremization, this picks up the extremal surface $\p_{\p \overline{C}} \Phi_{A/B}=0$, on  which the dilaton value is equal to  the area of the horizon of  the relevant black hole.
The dominant saddle of the gravitational path integral  can be found by taking the  minimum value  among these four candidates $\{\Phi_{AL/BR}, \; \Phi_{AL/BL}, \; \Phi_{AR/BL},\Phi_{AR/BR} \}$.  
On the other hand the value of the denominator is independent from the choice of the wormhole, i.e., in the absence of the cut, their action values are always the same. 

By combining these, we conclude that in the high temperature limit $\beta \rightarrow 0$,  the entanglement entropy is given by the sum of the two smallest black hole entropies. In our case, it is given by 
 \be
 S(\rho_{A})= {\rm Min} [S_{AL}+ S_{BL}, \; S_{AL}+ S_{BR},\; S_{AR}+S_{BR},\;S_{AR}+S_{BL}].
 \ee

We can also construct dilaton profiles describing general glued geometries in CGHS model too. This construction is parallel to  the one in the previous sections  \ref{sec:EntFlat}. In this case, we consider the following dilaton profile
\begin{equation}
	\begin{aligned}
			\Phi_A(x^+,x^-)&=\phi_{0,A}+\frac{|\Lambda|}{4}\tan x^{+}\tan x^{-} - 4\pi  G X \; (x^{+} \tan x^{+} + x^{-} \tan x^{-}) \\
			& \hspace{8cm}+ D_A(\tan x^{+} + \tan x^{-} ) 
	\end{aligned} 
\end{equation}
for the universe $A$, and similarly for the universe $B$. This dilaton profile $\Phi_A$ corresponds to a black hole with the masses
\begin{equation}
	M_{AR} = \frac{\sqrt{|\Lambda|}}{16\pi G}\left( \phi_{0,A} -\frac{4}{|\Lambda|} \left( D_A - 2\pi^2  G X \right)^{2} \right), \quad M_{AL} = \frac{\sqrt{|\Lambda|}}{16\pi G}\left( \phi_{0,A }-\frac{4}{|\Lambda|} \left( D_A + 2\pi^2  G X \right)^{2} \right),
\end{equation}
and similarly for $\Phi_B$.

For these masses, by repeating the analysis of the section \ref{sec:EntFlat}, we get various dilaton profiles $\Phi_{AL/BR}, \; \Phi_{AL/BR}, \; \Phi_{AR/BL},\Phi_{AR/BL}$ as in the case of JT gravity. Namely, each dilaton profile can be constructed starting from the general solution \eqref{eq:flatAB} with the replacement $ \la T \ra_{\beta} \rightarrow \la T \ra_{\beta} +X$ with the boundary conditions analgous to \eqref{eq:boundaryCondiCGHS}.

\section{Conclusion and discussions}\label{sec:conclusion}

In this paper, we consider the entanglement entropy of states defined on two disjoint universes, by generalizing the argument of \cite{Balasubramanian:2021wgd}.

Let us interpret the result we obtained.  We started from the entangled state \eqref{eq:TFD stateonAB} on two disjoint gravitating universes A and B. Since there is a black hole on the universe B, this state can be regarded as an entangled state of the Hawking quanta in the universe A and black hole microstates in the universe B. This interpretation is similar to the  setup for the island formula, where one takes the bath universe to be non gravitating. In the latter case, although the number of the degrees of freedom in the non-gravitating universe is infinite,  the maximal entanglement that the system can accommodate is given by the entropy of the black hole in the universe B in accord with the structure of the total Hilbert space.

In the current setup where both of these universes are gravitating, this scenario gets modified.   Namely, as is clear from the formula \eqref{eq:islandg}, the gravitational dynamics completely changes the system. After the dominant saddle in \eqref{eq:islandg} is changed, it  becomes a system where the Hawking radiation is entangled with the black holes in $A/B$, whose detailed properties were studied in this paper.  Thus we speculate the interior region of the new black hole belongs to the entanglement wedge of the Hawking radiation.  Such a gluing happens due to the back reaction of  the stress tensor induced from the entanglement between two universes, A and B. This can be regarded as a concrete realization of ER=EPR \cite{Maldacena:2013xja}, which relates entanglement in presence of gravity to spatial wormholes.  

It would be interesting to study this phenomenon further, with emphasis on its microscopic origin. 
One consistent description of an evaporating black hole is in terms of a class of states of the form 
\be
|\Psi \ra = \sum_{\alpha, i} C_{\alpha,i} | \psi_{\alpha} \ra_{BH} \otimes  |i \ra_{R}
\label{eq:randomstate}
\ee
where $C_{\alpha,i}$ is a random matrix drawn from Gaussian ensemble, and $| \psi_{\alpha} \ra_{BH}$, 
$ |i \ra_{R}$ are orthonomal basis of the Hilbert space of black hole microstates $H_{BH}$, and the similar basis for  Hawking radiation $H_{R}$ respectively. This  point of view of evaporating black holes was first studied by Page \cite{Page:1993wv}. The randomness of the coefficient matrix is coming from the chaotic nature of the black hole dynamics. Averaging over these random matrices leads to a semi-classical description, but consistent with the principles of quantum theory.  For example the Page curve of the radiation entropy as well as the island formula follows from it. From this point of view, the island region in the black hole interior can be regarded as a region accommodating  random fluctuations in the entangled state \eqref{eq:randomstate}. See \cite{Iizuka:2021tut} for a recent discussion on this topic and its relation to  baby universes. 
Since the system we studied in this paper involves two such  black holes, it is natural to  expect that the microscopic description of the setup has to do with two such random matrices. Related discussions toward this direction can be found in  \cite{Nomura:2018kia, Anderson:2021vof}. 

\section*{Acknowledgement}

TU thanks Vijay Balasubramanian, Arjun Kar for useful discussions in the related projects. TU was supported by JSPS Grant-in-Aid for Young Scientists 19K14716 and MEXT KAKENHI Grant-in-Aid for Transformative Research Areas A “Extreme Universe” No 21H05184.

\appendix

\appendix

\section{ADM Mass formula}\label{app:ADM}

In this appendix, we give ADM mass formulae in AdS JT gravity and CGHS gravity. We also give the explicit ADM masses of general dilaton profiles in both cases.

\subsection{ADM mass formula in AdS JT gravity}

We consider the ADM  mass of  a dilaton profile in AdS JT gravity. The relation between dilaton profiles and the ADM black hole mass is discussed in e.g. \cite{Maldacena:2016upp,Harlow:2018tqv}.
To introduce the ADM mass formula, we need to specify boundary conditions of the metric and the dilaton on the asymptotic boundary. We assume that they are given by
\begin{gather}
	\Phi|_{bdy}=\frac{\phi_{b}}{\varepsilon}, \label{eq:bdyConDila}\\
	g_{uu}|_{bdy}=-\frac{1}{\varepsilon^2},\label{eq:bdyConMetr}
\end{gather}
where $\varepsilon$ is the cutoff, $\phi_b$ is the renormalized dilaton value, which is a constant at the  cutoff surface, and $u$ is a boundary time coordinate along the  cutoff surface. Also $|_{bdy}$ implies that we evaluate the expression at the  boundary.

Then the ADM mass formula associated with one side of the two AdS boundaries is given by
\begin{equation}
	M_{ADM}=-\f{\sqrt{-g_{uu}}}{8\pi G}(\partial_n \Phi -\Phi)|_{bdy}, \label{eq:ADM-JT}
\end{equation}
where $n$ is the (outward) normal vector to the boundary, $\partial_{n}$ denotes the normal derivative to the boundary.

For example, we consider a dilaton profile $\Phi$ satisfying the equations of motion \eqref{eq:dilgen} with vanishing stress energy tensor.
The general solution is given by
\begin{equation}
\begin{aligned}
		\Phi &= Q_{-1}\frac{\cos \tau}{\cos \mu} + Q_{0}\frac{\sin \tau}{\cos \mu}-Q_{1} \tan \mu ,
\end{aligned}
\end{equation}
where $Q_{-1},Q_{0},Q_{1}$ are constants.

For this general dilaton profile, the ADM black hole mass associated with one side of the AdS boundaries becomes
\begin{equation}
	\begin{aligned}
		M_{ADM} =\frac{(Q_{-1})^2 +(Q_{0})^2- (Q_{1})^2}{16\pi G\phi_b}.
	\end{aligned}\label{eq:ADMJTExplicit}
\end{equation} 
This expression is manifestly invariant under the SL($2,\mathbb{R}$) rotation, which is a symmetry of JT gravity.

\subsection{ADM mass formula in Flat CGHS gravity}

Next we consider the ADM black hole mass described by a dilaton profile in CGHS gravity. For CGHS gravity, the relation between dilaton profiles and the ADM black hole mass is discussed in e.g. \cite{Cadoni:1995dd,Mann:1992yv}. 

The ADM mass formula associated with one side of the two spatial  infinities is given by
\begin{equation}
    M_{ADM}=\frac{1}{16\pi G\sqrt{|\Lambda|}}\left(- (\nabla\Phi)^2+|\Lambda|\Phi \right)|_{\rm sp.inf.},
\label{eq:ADMCado}
\end{equation}
where $|_{\rm sp.inf.}$ implies that we evaluate the expression at a spatial infinity.

As in the case of JT gravity, we consider a dilaton profile $\Phi$ satisfying the equations of motion \eqref{eq:eomForDiagoCom} with vanishing stress energy tensor.
The general solution is given by
\begin{equation}
    \begin{aligned}
    \Phi&= D^0+\frac{|\Lambda|}{4}\tan x^{+}  \tan  x^{-} + D^{+}\tan x^{+} + D^{-}\tan x^{-} ,
    \end{aligned}\label{eq:generalDilatonProfile}
\end{equation}
where $D^{0,\pm}$ are constants.

This dilaton profile gives the ADM black hole mass associated with one side of the two spatial infinities
\begin{equation}
    \begin{aligned}
    M_{ADM}&=\frac{\sqrt{|\Lambda|}}{16\pi G} \left( D^{0} -\frac{4}{\sqrt{|\Lambda|}}D^{+}D^{-} \right) .
    \end{aligned}\label{eq:ADM_GeneralDila} 
\end{equation}

\section{The black hole solution  with a shockwave in its interior (AdS JT gravity)} \label{app:ShockJT}

In this appendix, we explain a method to construct a dilaton profile describing a glued spacetime approximately by using  shock waves for AdS JT gravity  outlined in rection \ref{sec:apglue}. 
 This  is closely related to the exact construction in section \ref{sec:EntADS}, but the resulting geometry does not have the time reflection symmetry $\tau \longleftrightarrow -\tau$\footnote{We can construct a dilaton profile describing a glued geometry which is not time reflection symmetric with the method discussed in section \ref{sec:EntADS} by including the additional term in \eqref{eq:dilgen}. See footnote \cref{foot:JTAddito}.}.

At first, we consider the equations of motion \eqref{eq:dilgen}
with the stress tensor given by
\be
\la T_{\pm \pm} \ra = \la T_{\pm \pm} \ra_{\beta} +\la T_{\pm \pm} \ra_{S},\quad \la T_{\pm \mp} \ra = 0,\label{eq:StresEnergyTensorWiShock}
\ee
where  the first term $\la T_{\pm \pm} \ra_{\beta}$ is defined by \eqref{eq:SEtensor} and the second term $\la T_{\pm \pm} \ra_{S}$ is coming from the contribution of the shock wave, 
\be
\la T_{+ +} \ra_{S} =E  \delta (x^{+}-x^{+}_{0}),\quad \la T_{- -} \ra_{S} =0.
\ee
Here, the coefficient $E$ characterizes the strength of the shock wave and $x_0^{+}$ is the location of the shock wave.

The general solution is given by\footnote{Although the full general solution includes terms $\sin\tau/\cos\mu$ and $\tan \mu$, we omit such terms for simplicity.}
\begin{align}
\Phi &= b_{+} \left(\f{\cos \tau}{\cos \mu} \right) -16\pi G \la T \ra_{\beta} ( \mu \tan \mu +1) \nonumber \\[+10pt]
&\qquad - 16 \pi G E  
\cos^{2} \left(\f{x^{+}_{0} + x^{-}}{2}\right) \left[ \tan \mu -\tan \left(\f{x^{+}_{0} + x^{-}}{2} \right) \right] \Theta (x^{+} -x^{+}_{0} ).\label{eq:fullAdSdilaton}
\end{align}

As we will check in the next subsection, this dilaton profile corresponds to a black hole with two different masses. In other words, this dilaton profile describes a new glued spacetime with its masses.

\subsection{Black hole ADM Mass formula in the AdS Jackiw-Teitelboim gravity}
Next we compute the black hole masses described by the dilaton profile \eqref{eq:fullAdSdilaton} by using the ADM mass formula \eqref{eq:ADM-JT}. 

To use the ADM black hole mass formula, we need to calculate the normal derivative of the dilaton $\partial_n \Phi$. After some algebras, we get
\begin{equation}
	\begin{aligned}
		\partial_n \Phi 
		&=\Bigg[\Phi^2  - \bigg\{
		b_{+}\left(b_{+} + 16 \pi G E \sin x_{0}^{+} \Theta\left(x^{+}-x_{0}^{+}\right) \right)
		 -16\pi G \la T \ra_{\beta}  \cdot 16 \pi G E \cdot \mu\, \Theta\left(x^{+}-x_{0}^{+}\right) \\
		& \hspace{11.5cm} - \left(16\pi G \la T \ra_{\beta}  \right)^2 \cdot \mu^{2}  \bigg\}\\
		&\hspace{3cm} +16\pi G \la T \ra_{\beta} \cos \mu \bigg\{2 (b_{+}+8 \pi G E \sin x_{0}^{+} \Theta\left(x^{+}-x_{0}^{+}\right))\cos t\\
		& \hspace{7cm} - 16 \pi G E \cos x_{0} ^{+} \sin t - 16\pi G \la T \ra_{\beta}  \cos \mu \bigg\}+\cdots \Bigg]^{\frac{1}{2}},
	\end{aligned}
\end{equation}
where the dots $\cdots$ denotes terms containing the delta function, which comes from the derivative of the step function $\Theta\left(x^{+}-x_{0}^{+}\right)$, and the terms do not contribute to the black hole mass and below we ignore such terms.

Therefore, from the ADM black hole mass formula \eqref{eq:ADM-JT} and the boundary conditions \eqref{eq:bdyConDila}, \eqref{eq:bdyConMetr}, the black hole masses associated with two AdS boundaries are 
\begin{equation}
	\begin{aligned}
			M_{ADM,\text{L}}&=-\f{\sqrt{-g_{uu}}}{8\pi G}(\partial_n \Phi -\Phi)|_{\text{Left }bdy,\; \mu \to -\pi/2}\\
		&=\frac{1}{16\pi G\phi_b} \bigg\{
		b_{+}^{2} - \left( 8\pi^2 G \la T \ra_{\beta}  \right)^2  \bigg\}
	\end{aligned}
\end{equation} 
and 
\begin{equation}
	\begin{aligned}
			M_{ADM,\text{R}}&=-\f{\sqrt{-g_{uu}}}{8\pi G}(\partial_n \Phi -\Phi)|_{\text{Right }bdy,\; \mu \to \pi/2}\\
		&=\frac{1}{16\pi G\phi_b}  \bigg\{
		b_{+}\left(b_{+}+16 \pi G E \sin x_{0}^{+}\right) -      128 \pi^2 G^2 E \la T \ra_{\beta} - \left( 8\pi^2 G \la T \ra_{\beta}  \right)^2  \bigg\}.
	\end{aligned}
\end{equation}

Since these ADS black hole masses associated with the dilaton profile \eqref{eq:fullAdSdilaton}
are clearly different $M_{ADM,\text{L}}\neq M_{ADM,\text{R}}$ unless $E=0$, thus the dilaton profile \eqref{eq:fullAdSdilaton} describes the glued geometry with different black hole masses $M_{ADM,\text{L}},M_{ADM,\text{R}}$.

\section{The black hole solution  with a shockwave in its interior (Flat CGHS gravity)} \label{app:ShockCGHS}

In this appendix, for CGHS gravity in flat spacetime, we explain a method to construct a dilaton profile describing a glued spacetime approximately by using a shock wave.
 The discussion is parallel to the JT gravity case. This method is also closely related to the one in section \ref{sec:EntFlat}, and the resulting geometry also does not have the time reflection symmetry $t \longleftrightarrow -t$ as in the case of JT gravity\footnote{In this case, we can also construct a glued geometry which is not time reflection symmetric with the method discussed in section \ref{sec:EntFlat} by considering $D^{+}\neq D^{-}$ cases.}.

Firstly, we consider the equations of motion \eqref{eq:eomForDiagoCom}
with the stress energy tensor given by the previous one \eqref{eq:StresEnergyTensorWiShock}. 

The full solution is given by\footnote{The full general solution also includes terms $\tan x^{+} $ and $\tan x^{-}$, but we omit them for simplicity.}
\begin{equation}
\begin{aligned}
    \Phi&=\phi_{0}+\frac{|\Lambda|}{4} \tan x^{+} \tan x^{-} -4\pi G \la T \ra_{\beta}(x^+ \tan x^+ + x^- \tan x^-)\\
    &\hspace{5cm}  - 8\pi G E\cos^2x_0^+\left( \tan x^+ -\tan x_0^+ \right)\Theta(x^+ - x_0^+).
\end{aligned}\label{eq:dilatonProfSourceShock}
\end{equation}

This dilaton profile \eqref{eq:dilatonProfSourceShock} corresponds to a black hole with two different masses in CGHS gravity. In other words, this dilaton profile describes a new glued spacetime with its masses.

\subsection{Black hole ADM Mass formula in CGHS gravity}

Next we compute the black hole mass described by the dilaton profile \eqref{eq:dilatonProfSourceShock} by using the ADM mass formula \eqref{eq:ADMCado}. 

To use the ADM mass formula \eqref{eq:ADMCado}, we need to evaluate the factor $(\nabla\Phi)^2$, which is given by 
\begin{equation}
	\begin{aligned}
	(\nabla\Phi)^2
    &=4 \left[\frac{|\Lambda|}{4}\tan x^{-} -4\pi G \la T \ra_{\beta}(\cos x^{+}\sin x^{+} + x^{+} ) \right]\\
    &\hspace{3cm}\times \left[\frac{|\Lambda|}{4}\tan x^{+} - 4\pi G \la T \ra_{\beta}(\cos x^{-}\sin x^{-} + x^{-} ) \right]	\\
    &\hspace{4cm} - 32\pi G E \cos^{2} x_{0}^{+} \left[\frac{|\Lambda|}{4}\tan x^{+} - 4\pi G \la T \ra_{\beta} (\cos x^{-}\sin x^{-} + x^{-} )
    \right].
    \end{aligned}
\end{equation}

Thus the black hole masses at the left and right asymptotic spatial infinity are given by
\begin{equation}
	\begin{aligned}
		M_{ADM,\text{L}}&=\lim_{x^{\pm}\to -\frac{\pi}{2} } \frac{1}{16\pi G \sqrt{|\Lambda|}}\left(-(\nabla \Phi)^{2}+|\Lambda| \Phi\right)	\\
		&= \frac{\sqrt{|\Lambda|}}{16\pi G } \bigg\{ \phi_{0} -\frac{ \left(4\pi^2 G \la T \ra_{\beta}\right)^{2} }{|\Lambda|} \Bigg\},
	\end{aligned}
\end{equation}
and
\begin{equation}
	\begin{aligned}
		M_{ADM,\text{R}}&=\lim_{x^{\pm}\to \frac{\pi}{2} } \frac{1}{16\pi G \sqrt{|\Lambda|}}\left(-(\nabla \Phi)^{2}+|\Lambda| \Phi\right)	\\
		&= \frac{\sqrt{|\Lambda|}}{16\pi G } \bigg\{ \phi_{0} -\frac{ \left(4\pi^2 G \la T \ra_{\beta}\right)^{2} }{|\Lambda|}-\frac{64 \pi^3 G^2 E \la T \ra_{\beta} \cos ^{2} x_{0}^{+}}{|\Lambda|} +8 \pi G E \cos x_{0}^{+} \sin x_{0}^{+}  \Bigg\}.
	\end{aligned}
\end{equation}

In this case, since the black hole masses $M_{ADM,\text{L}},M_{ADM,\text{R}}$ are clearly different again, the dilaton profile \eqref{eq:dilatonProfSourceShock} describes the black hole with two different black hole masses.

\bibliographystyle{JHEP}
\bibliography{island}

\providecommand{\href}[2]{#2}\begingroup\raggedright\begin{thebibliography}{10}

\bibitem{Hawking:1975vcx}
S.~W. Hawking, \emph{{Particle Creation by Black Holes}},
  \href{https://doi.org/10.1007/BF02345020}{\emph{Commun. Math. Phys.}
  {\bfseries 43} (1975) 199}.

\bibitem{Hawking:1976ra}
S.~W. Hawking, \emph{{Breakdown of Predictability in Gravitational Collapse}},
  \href{https://doi.org/10.1103/PhysRevD.14.2460}{\emph{Phys. Rev. D}
  {\bfseries 14} (1976) 2460}.

\bibitem{Page:1993wv}
D.~N. Page, \emph{{Information in black hole radiation}},
  \href{https://doi.org/10.1103/PhysRevLett.71.3743}{\emph{Phys. Rev. Lett.}
  {\bfseries 71} (1993) 3743}
  [\href{https://arxiv.org/abs/hep-th/9306083}{{\ttfamily hep-th/9306083}}].

\bibitem{Page:2013dx}
D.~N. Page, \emph{{Time Dependence of Hawking Radiation Entropy}},
  \href{https://doi.org/10.1088/1475-7516/2013/09/028}{\emph{JCAP} {\bfseries
  09} (2013) 028} [\href{https://arxiv.org/abs/1301.4995}{{\ttfamily
  1301.4995}}].

\bibitem{Penington:2019npb}
G.~Penington, \emph{{Entanglement Wedge Reconstruction and the Information
  Paradox}},  \href{https://arxiv.org/abs/1905.08255}{{\ttfamily 1905.08255}}.

\bibitem{Almheiri:2019psf}
A.~Almheiri, N.~Engelhardt, D.~Marolf and H.~Maxfield, \emph{{The entropy of
  bulk quantum fields and the entanglement wedge of an evaporating black
  hole}}, \href{https://doi.org/10.1007/JHEP12(2019)063}{\emph{JHEP} {\bfseries
  12} (2019) 063} [\href{https://arxiv.org/abs/1905.08762}{{\ttfamily
  1905.08762}}].

\bibitem{Almheiri:2019hni}
A.~Almheiri, R.~Mahajan, J.~Maldacena and Y.~Zhao, \emph{{The Page curve of
  Hawking radiation from semiclassical geometry}},
  \href{https://doi.org/10.1007/JHEP03(2020)149}{\emph{JHEP} {\bfseries 03}
  (2020) 149} [\href{https://arxiv.org/abs/1908.10996}{{\ttfamily
  1908.10996}}].

\bibitem{Penington:2019kki}
G.~Penington, S.~H. Shenker, D.~Stanford and Z.~Yang, \emph{{Replica wormholes
  and the black hole interior}},
  \href{https://arxiv.org/abs/1911.11977}{{\ttfamily 1911.11977}}.

\bibitem{Almheiri:2019qdq}
A.~Almheiri, T.~Hartman, J.~Maldacena, E.~Shaghoulian and A.~Tajdini,
  \emph{{Replica Wormholes and the Entropy of Hawking Radiation}},
  \href{https://doi.org/10.1007/JHEP05(2020)013}{\emph{JHEP} {\bfseries 05}
  (2020) 013} [\href{https://arxiv.org/abs/1911.12333}{{\ttfamily
  1911.12333}}].

\bibitem{Rozali:2019day}
M.~Rozali, J.~Sully, M.~Van~Raamsdonk, C.~Waddell and D.~Wakeham,
  \emph{{Information radiation in BCFT models of black holes}},
  \href{https://arxiv.org/abs/1910.12836}{{\ttfamily 1910.12836}}.

\bibitem{Chen:2019uhq}
H.~Z. Chen, Z.~Fisher, J.~Hernandez, R.~C. Myers and S.-M. Ruan,
  \emph{{Information Flow in Black Hole Evaporation}},
  \href{https://doi.org/10.1007/JHEP03(2020)152}{\emph{JHEP} {\bfseries 03}
  (2020) 152} [\href{https://arxiv.org/abs/1911.03402}{{\ttfamily
  1911.03402}}].

\bibitem{Bousso:2019ykv}
R.~Bousso and M.~Toma\v{s}evi\'c, \emph{{Unitarity From a Smooth Horizon?}},
  \href{https://doi.org/10.1103/PhysRevD.102.106019}{\emph{Phys. Rev. D}
  {\bfseries 102} (2020) 106019}
  [\href{https://arxiv.org/abs/1911.06305}{{\ttfamily 1911.06305}}].

\bibitem{Almheiri:2019psy}
A.~Almheiri, R.~Mahajan and J.~E. Santos, \emph{{Entanglement islands in higher
  dimensions}},  \href{https://arxiv.org/abs/1911.09666}{{\ttfamily
  1911.09666}}.

\bibitem{Chen:2019iro}
Y.~Chen, \emph{{Pulling Out the Island with Modular Flow}},
  \href{https://arxiv.org/abs/1912.02210}{{\ttfamily 1912.02210}}.

\bibitem{Balasubramanian:2020hfs}
V.~Balasubramanian, A.~Kar, O.~Parrikar, G.~Sárosi and T.~Ugajin,
  \emph{{Geometric secret sharing in a model of Hawking radiation}},
  \href{https://arxiv.org/abs/2003.05448}{{\ttfamily 2003.05448}}.

\bibitem{Hollowood:2020cou}
T.~J. Hollowood and S.~P. Kumar, \emph{{Islands and Page Curves for Evaporating
  Black Holes in JT Gravity}},
  \href{https://arxiv.org/abs/2004.14944}{{\ttfamily 2004.14944}}.

\bibitem{Alishahiha:2020qza}
M.~Alishahiha, A.~Faraji~Astaneh and A.~Naseh, \emph{{Island in the presence of
  higher derivative terms}},
  \href{https://doi.org/10.1007/JHEP02(2021)035}{\emph{JHEP} {\bfseries 02}
  (2021) 035} [\href{https://arxiv.org/abs/2005.08715}{{\ttfamily
  2005.08715}}].

\bibitem{Chen:2020uac}
H.~Z. Chen, R.~C. Myers, D.~Neuenfeld, I.~A. Reyes and J.~Sandor,
  \emph{{Quantum Extremal Islands Made Easy, Part I: Entanglement on the
  Brane}},  \href{https://arxiv.org/abs/2006.04851}{{\ttfamily 2006.04851}}.

\bibitem{Geng:2020qvw}
H.~Geng and A.~Karch, \emph{{Massive Islands}},
  \href{https://arxiv.org/abs/2006.02438}{{\ttfamily 2006.02438}}.

\bibitem{Chandrasekaran:2020qtn}
V.~Chandrasekaran, M.~Miyaji and P.~Rath, \emph{{Including contributions from
  entanglement islands to the reflected entropy}},
  \href{https://doi.org/10.1103/PhysRevD.102.086009}{\emph{Phys. Rev. D}
  {\bfseries 102} (2020) 086009}
  [\href{https://arxiv.org/abs/2006.10754}{{\ttfamily 2006.10754}}].

\bibitem{Li:2020ceg}
T.~Li, J.~Chu and Y.~Zhou, \emph{{Reflected Entropy for an Evaporating Black
  Hole}}, \href{https://doi.org/10.1007/JHEP11(2020)155}{\emph{JHEP} {\bfseries
  11} (2020) 155} [\href{https://arxiv.org/abs/2006.10846}{{\ttfamily
  2006.10846}}].

\bibitem{Bousso:2020kmy}
R.~Bousso and E.~Wildenhain, \emph{{Gravity/ensemble duality}},
  \href{https://doi.org/10.1103/PhysRevD.102.066005}{\emph{Phys. Rev. D}
  {\bfseries 102} (2020) 066005}
  [\href{https://arxiv.org/abs/2006.16289}{{\ttfamily 2006.16289}}].

\bibitem{Dong:2020uxp}
X.~Dong, X.-L. Qi, Z.~Shangnan and Z.~Yang, \emph{{Effective entropy of quantum
  fields coupled with gravity}},
  \href{https://arxiv.org/abs/2007.02987}{{\ttfamily 2007.02987}}.

\bibitem{Hollowood:2020kvk}
T.~J. Hollowood, S.~Prem~Kumar and A.~Legramandi, \emph{{Hawking radiation
  correlations of evaporating black holes in JT gravity}},
  \href{https://doi.org/10.1088/1751-8121/abbc51}{\emph{J. Phys. A} {\bfseries
  53} (2020) 475401} [\href{https://arxiv.org/abs/2007.04877}{{\ttfamily
  2007.04877}}].

\bibitem{Chen:2020jvn}
H.~Z. Chen, Z.~Fisher, J.~Hernandez, R.~C. Myers and S.-M. Ruan,
  \emph{{Evaporating Black Holes Coupled to a Thermal Bath}},
  \href{https://doi.org/10.1007/JHEP01(2021)065}{\emph{JHEP} {\bfseries 01}
  (2021) 065} [\href{https://arxiv.org/abs/2007.11658}{{\ttfamily
  2007.11658}}].

\bibitem{Chen:2020tes}
Y.~Chen, V.~Gorbenko and J.~Maldacena, \emph{{Bra-ket wormholes in
  gravitationally prepared states}},
  \href{https://arxiv.org/abs/2007.16091}{{\ttfamily 2007.16091}}.

\bibitem{Hartman:2020khs}
T.~Hartman, Y.~Jiang and E.~Shaghoulian, \emph{{Islands in cosmology}},
  \href{https://arxiv.org/abs/2008.01022}{{\ttfamily 2008.01022}}.

\bibitem{Balasubramanian:2020coy}
V.~Balasubramanian, A.~Kar and T.~Ugajin, \emph{{Entanglement between two
  disjoint universes}},
  \href{https://doi.org/10.1007/JHEP02(2021)136}{\emph{JHEP} {\bfseries 02}
  (2021) 136} [\href{https://arxiv.org/abs/2008.05274}{{\ttfamily
  2008.05274}}].

\bibitem{Balasubramanian:2020xqf}
V.~Balasubramanian, A.~Kar and T.~Ugajin, \emph{{Islands in de Sitter space}},
  \href{https://doi.org/10.1007/JHEP02(2021)072}{\emph{JHEP} {\bfseries 02}
  (2021) 072} [\href{https://arxiv.org/abs/2008.05275}{{\ttfamily
  2008.05275}}].

\bibitem{Ling:2020laa}
Y.~Ling, Y.~Liu and Z.-Y. Xian, \emph{{Island in Charged Black Holes}},
  \href{https://doi.org/10.1007/JHEP03(2021)251}{\emph{JHEP} {\bfseries 03}
  (2021) 251} [\href{https://arxiv.org/abs/2010.00037}{{\ttfamily
  2010.00037}}].

\bibitem{Chen:2020hmv}
H.~Z. Chen, R.~C. Myers, D.~Neuenfeld, I.~A. Reyes and J.~Sandor,
  \emph{{Quantum Extremal Islands Made Easy, Part II: Black Holes on the
  Brane}}, \href{https://doi.org/10.1007/JHEP12(2020)025}{\emph{JHEP}
  {\bfseries 12} (2020) 025}
  [\href{https://arxiv.org/abs/2010.00018}{{\ttfamily 2010.00018}}].

\bibitem{Bhattacharya:2020uun}
A.~Bhattacharya, A.~Chanda, S.~Maulik, C.~Northe and S.~Roy, \emph{{Topological
  shadows and complexity of islands in multiboundary wormholes}},
  \href{https://doi.org/10.1007/JHEP02(2021)152}{\emph{JHEP} {\bfseries 02}
  (2021) 152} [\href{https://arxiv.org/abs/2010.04134}{{\ttfamily
  2010.04134}}].

\bibitem{Harlow:2020bee}
D.~Harlow and E.~Shaghoulian, \emph{{Global symmetry, Euclidean gravity, and
  the black hole information problem}},
  \href{https://doi.org/10.1007/JHEP04(2021)175}{\emph{JHEP} {\bfseries 04}
  (2021) 175} [\href{https://arxiv.org/abs/2010.10539}{{\ttfamily
  2010.10539}}].

\bibitem{Akal:2020ujg}
I.~Akal, \emph{{Universality, intertwiners and black hole information}},
  \href{https://arxiv.org/abs/2010.12565}{{\ttfamily 2010.12565}}.

\bibitem{Hernandez:2020nem}
J.~Hernandez, R.~C. Myers and S.-M. Ruan, \emph{{Quantum extremal islands made
  easy. Part III. Complexity on the brane}},
  \href{https://doi.org/10.1007/JHEP02(2021)173}{\emph{JHEP} {\bfseries 02}
  (2021) 173} [\href{https://arxiv.org/abs/2010.16398}{{\ttfamily
  2010.16398}}].

\bibitem{Matsuo:2020ypv}
Y.~Matsuo, \emph{{Islands and stretched horizon}},
  \href{https://arxiv.org/abs/2011.08814}{{\ttfamily 2011.08814}}.

\bibitem{Akal:2020twv}
I.~Akal, Y.~Kusuki, N.~Shiba, T.~Takayanagi and Z.~Wei, \emph{{Entanglement
  Entropy in a Holographic Moving Mirror and the Page Curve}},
  \href{https://doi.org/10.1103/PhysRevLett.126.061604}{\emph{Phys. Rev. Lett.}
  {\bfseries 126} (2021) 061604}
  [\href{https://arxiv.org/abs/2011.12005}{{\ttfamily 2011.12005}}].

\bibitem{Numasawa:2020sty}
T.~Numasawa, \emph{{Four coupled SYK models and Nearly AdS$_2$ gravities: Phase
  Transitions in Traversable wormholes and in Bra-ket wormholes}},
  \href{https://arxiv.org/abs/2011.12962}{{\ttfamily 2011.12962}}.

\bibitem{KumarBasak:2020ams}
J.~Kumar~Basak, D.~Basu, V.~Malvimat, H.~Parihar and G.~Sengupta,
  \emph{{Islands for Entanglement Negativity}},
  \href{https://arxiv.org/abs/2012.03983}{{\ttfamily 2012.03983}}.

\bibitem{Geng:2020fxl}
H.~Geng, A.~Karch, C.~Perez-Pardavila, S.~Raju, L.~Randall, M.~Riojas et~al.,
  \emph{{Information Transfer with a Gravitating Bath}},
  \href{https://arxiv.org/abs/2012.04671}{{\ttfamily 2012.04671}}.

\bibitem{Deng:2020ent}
F.~Deng, J.~Chu and Y.~Zhou, \emph{{Defect extremal surface as the holographic
  counterpart of Island formula}},
  \href{https://doi.org/10.1007/JHEP03(2021)008}{\emph{JHEP} {\bfseries 03}
  (2021) 008} [\href{https://arxiv.org/abs/2012.07612}{{\ttfamily
  2012.07612}}].

\bibitem{Karananas:2020fwx}
G.~K. Karananas, A.~Kehagias and J.~Taskas, \emph{{Islands in Linear Dilaton
  Black Holes}},  \href{https://arxiv.org/abs/2101.00024}{{\ttfamily
  2101.00024}}.

\bibitem{Wang:2021woy}
X.~Wang, R.~Li and J.~Wang, \emph{{Islands and Page curves of
  Reissner-Nordstr\"om black holes}},
  \href{https://doi.org/10.1007/JHEP04(2021)103}{\emph{JHEP} {\bfseries 04}
  (2021) 103} [\href{https://arxiv.org/abs/2101.06867}{{\ttfamily
  2101.06867}}].

\bibitem{Kawabata:2021hac}
K.~Kawabata, T.~Nishioka, Y.~Okuyama and K.~Watanabe, \emph{{Probing Hawking
  radiation through capacity of entanglement}},
  \href{https://arxiv.org/abs/2102.02425}{{\ttfamily 2102.02425}}.

\bibitem{Fallows:2021sge}
S.~Fallows and S.~F. Ross, \emph{{Islands and mixed states in closed
  universes}},  \href{https://arxiv.org/abs/2103.14364}{{\ttfamily
  2103.14364}}.

\bibitem{Bhattacharya:2021jrn}
A.~Bhattacharya, A.~Bhattacharyya, P.~Nandy and A.~K. Patra, \emph{{Islands and
  complexity of eternal black hole and radiation subsystems for a doubly
  holographic model}},
  \href{https://doi.org/10.1007/JHEP05(2021)135}{\emph{JHEP} {\bfseries 05}
  (2021) 135} [\href{https://arxiv.org/abs/2103.15852}{{\ttfamily
  2103.15852}}].

\bibitem{Kim:2021gzd}
W.~Kim and M.~Nam, \emph{{Entanglement entropy of asymptotically flat
  non-extremal and extremal black holes with an island}},
  \href{https://doi.org/10.1140/epjc/s10052-021-09680-x}{\emph{Eur. Phys. J. C}
  {\bfseries 81} (2021) 869}
  [\href{https://arxiv.org/abs/2103.16163}{{\ttfamily 2103.16163}}].

\bibitem{Anderson:2021vof}
L.~Anderson, O.~Parrikar and R.~M. Soni, \emph{{Islands with Gravitating
  Baths}},  \href{https://arxiv.org/abs/2103.14746}{{\ttfamily 2103.14746}}.

\bibitem{Miyata:2021ncm}
A.~Miyata and T.~Ugajin, \emph{{Evaporation of black holes in flat space
  entangled with an auxiliary universe}},
  \href{https://arxiv.org/abs/2104.00183}{{\ttfamily 2104.00183}}.

\bibitem{Wang:2021mqq}
X.~Wang, R.~Li and J.~Wang, \emph{{Islands and Page curves for a family of
  exactly solvable evaporating black holes}},
  \href{https://arxiv.org/abs/2104.00224}{{\ttfamily 2104.00224}}.

\bibitem{Ghosh:2021axl}
K.~Ghosh and C.~Krishnan, \emph{{Dirichlet baths and the not-so-fine-grained
  Page curve}}, \href{https://doi.org/10.1007/JHEP08(2021)119}{\emph{JHEP}
  {\bfseries 08} (2021) 119}
  [\href{https://arxiv.org/abs/2103.17253}{{\ttfamily 2103.17253}}].

\bibitem{Aalsma:2021bit}
L.~Aalsma and W.~Sybesma, \emph{{The Price of Curiosity: Information Recovery
  in de Sitter Space}},  \href{https://arxiv.org/abs/2104.00006}{{\ttfamily
  2104.00006}}.

\bibitem{Geng:2021iyq}
H.~Geng, S.~L\"ust, R.~K. Mishra and D.~Wakeham, \emph{{Holographic BCFTs and
  Communicating Black Holes}},
  \href{https://arxiv.org/abs/2104.07039}{{\ttfamily 2104.07039}}.

\bibitem{Balasubramanian:2021wgd}
V.~Balasubramanian, A.~Kar and T.~Ugajin, \emph{{Entanglement between two
  gravitating universes}},  \href{https://arxiv.org/abs/2104.13383}{{\ttfamily
  2104.13383}}.

\bibitem{Uhlemann:2021nhu}
C.~F. Uhlemann, \emph{{Islands and Page curves in 4d from Type IIB}},
  \href{https://doi.org/10.1007/JHEP08(2021)104}{\emph{JHEP} {\bfseries 08}
  (2021) 104} [\href{https://arxiv.org/abs/2105.00008}{{\ttfamily
  2105.00008}}].

\bibitem{Qi:2021sxb}
X.-L. Qi, \emph{{Entanglement island, miracle operators and the firewall}},
  \href{https://arxiv.org/abs/2105.06579}{{\ttfamily 2105.06579}}.

\bibitem{Kawabata:2021vyo}
K.~Kawabata, T.~Nishioka, Y.~Okuyama and K.~Watanabe, \emph{{Replica wormholes
  and capacity of entanglement}},
  \href{https://doi.org/10.1007/JHEP10(2021)227}{\emph{JHEP} {\bfseries 10}
  (2021) 227} [\href{https://arxiv.org/abs/2105.08396}{{\ttfamily
  2105.08396}}].

\bibitem{Chu:2021gdb}
J.~Chu, F.~Deng and Y.~Zhou, \emph{{Page curve from defect extremal surface and
  island in higher dimensions}},
  \href{https://doi.org/10.1007/JHEP10(2021)149}{\emph{JHEP} {\bfseries 10}
  (2021) 149} [\href{https://arxiv.org/abs/2105.09106}{{\ttfamily
  2105.09106}}].

\bibitem{Langhoff:2021uct}
K.~Langhoff, C.~Murdia and Y.~Nomura, \emph{{Multiverse in an inverted
  island}}, \href{https://doi.org/10.1103/PhysRevD.104.086007}{\emph{Phys. Rev.
  D} {\bfseries 104} (2021) 086007}
  [\href{https://arxiv.org/abs/2106.05271}{{\ttfamily 2106.05271}}].

\bibitem{Lu:2021gmv}
Y.~Lu and J.~Lin, \emph{{Islands in Kaluza-Klein black holes}},
  \href{https://arxiv.org/abs/2106.07845}{{\ttfamily 2106.07845}}.

\bibitem{Akal:2021foz}
I.~Akal, Y.~Kusuki, N.~Shiba, T.~Takayanagi and Z.~Wei, \emph{{Holographic
  moving mirrors}},
  \href{https://doi.org/10.1088/1361-6382/ac2c1b}{\emph{Class. Quant. Grav.}
  {\bfseries 38} (2021) 224001}
  [\href{https://arxiv.org/abs/2106.11179}{{\ttfamily 2106.11179}}].

\bibitem{Balasubramanian:2021xcm}
V.~Balasubramanian, B.~Craps, M.~Khramtsov and E.~Shaghoulian,
  \emph{{Submerging islands through thermalization}},
  \href{https://doi.org/10.1007/JHEP10(2021)048}{\emph{JHEP} {\bfseries 10}
  (2021) 048} [\href{https://arxiv.org/abs/2107.14746}{{\ttfamily
  2107.14746}}].

\bibitem{Ahn:2021chg}
B.~Ahn, S.-E. Bak, H.-S. Jeong, K.-Y. Kim and Y.-W. Sun, \emph{{Islands in
  charged linear dilaton black holes}},
  \href{https://arxiv.org/abs/2107.07444}{{\ttfamily 2107.07444}}.

\bibitem{Miyaji:2021lcq}
M.~Miyaji, \emph{{Island for Gravitationally Prepared State and Pseudo
  Entanglement Wedge}},  \href{https://arxiv.org/abs/2109.03830}{{\ttfamily
  2109.03830}}.

\bibitem{Matsuo:2021mmi}
Y.~Matsuo, \emph{{Entanglement entropy and vacuum states in Schwarzschild
  geometry}},  \href{https://arxiv.org/abs/2110.13898}{{\ttfamily 2110.13898}}.

\bibitem{Chen:2020ojn}
Y.~Chen and H.~W. Lin, \emph{{Signatures of global symmetry violation in
  relative entropies and replica wormholes}},
  \href{https://doi.org/10.1007/JHEP03(2021)040}{\emph{JHEP} {\bfseries 03}
  (2021) 040} [\href{https://arxiv.org/abs/2011.06005}{{\ttfamily
  2011.06005}}].

\bibitem{Hsin:2020mfa}
P.-S. Hsin, L.~V. Iliesiu and Z.~Yang, \emph{{A violation of global symmetries
  from replica wormholes and the fate of black hole remnants}},
  \href{https://arxiv.org/abs/2011.09444}{{\ttfamily 2011.09444}}.

\bibitem{Engelhardt:2020qpv}
N.~Engelhardt, S.~Fischetti and A.~Maloney, \emph{{Free energy from replica
  wormholes}}, \href{https://doi.org/10.1103/PhysRevD.103.046021}{\emph{Phys.
  Rev. D} {\bfseries 103} (2021) 046021}
  [\href{https://arxiv.org/abs/2007.07444}{{\ttfamily 2007.07444}}].

\bibitem{Karlsson:2020uga}
A.~Karlsson, \emph{{Replica wormhole and island incompatibility with monogamy
  of entanglement}},  \href{https://arxiv.org/abs/2007.10523}{{\ttfamily
  2007.10523}}.

\bibitem{Goto:2020wnk}
K.~Goto, T.~Hartman and A.~Tajdini, \emph{{Replica wormholes for an evaporating
  2D black hole}},  \href{https://arxiv.org/abs/2011.09043}{{\ttfamily
  2011.09043}}.

\bibitem{Hirano:2021rzg}
S.~Hirano and T.~Kuroki, \emph{{Replica Wormholes from Liouville Theory}},
  \href{https://arxiv.org/abs/2109.12539}{{\ttfamily 2109.12539}}.

\bibitem{Dong:2021oad}
X.~Dong, S.~McBride and W.~W. Weng, \emph{{Replica Wormholes and Holographic
  Entanglement Negativity}},
  \href{https://arxiv.org/abs/2110.11947}{{\ttfamily 2110.11947}}.

\bibitem{Almheiri:2019yqk}
A.~Almheiri, R.~Mahajan and J.~Maldacena, \emph{{Islands outside the horizon}},
   \href{https://arxiv.org/abs/1910.11077}{{\ttfamily 1910.11077}}.

\bibitem{Donnelly:2016auv}
W.~Donnelly and L.~Freidel, \emph{{Local subsystems in gauge theory and
  gravity}}, \href{https://doi.org/10.1007/JHEP09(2016)102}{\emph{JHEP}
  {\bfseries 09} (2016) 102}
  [\href{https://arxiv.org/abs/1601.04744}{{\ttfamily 1601.04744}}].

\bibitem{Raju:2021lwh}
S.~Raju, \emph{{Failure of the split property in gravity and the information
  paradox}},  \href{https://arxiv.org/abs/2110.05470}{{\ttfamily 2110.05470}}.

\bibitem{Maldacena:2013xja}
J.~Maldacena and L.~Susskind, \emph{{Cool horizons for entangled black holes}},
  \href{https://doi.org/10.1002/prop.201300020}{\emph{Fortsch. Phys.}
  {\bfseries 61} (2013) 781} [\href{https://arxiv.org/abs/1306.0533}{{\ttfamily
  1306.0533}}].

\bibitem{Lewkowycz:2013nqa}
A.~Lewkowycz and J.~Maldacena, \emph{{Generalized gravitational entropy}},
  \href{https://doi.org/10.1007/JHEP08(2013)090}{\emph{JHEP} {\bfseries 08}
  (2013) 090} [\href{https://arxiv.org/abs/1304.4926}{{\ttfamily 1304.4926}}].

\bibitem{Goel:2018ubv}
A.~Goel, H.~T. Lam, G.~J. Turiaci and H.~Verlinde, \emph{{Expanding the Black
  Hole Interior: Partially Entangled Thermal States in SYK}},
  \href{https://doi.org/10.1007/JHEP02(2019)156}{\emph{JHEP} {\bfseries 02}
  (2019) 156} [\href{https://arxiv.org/abs/1807.03916}{{\ttfamily
  1807.03916}}].

\bibitem{Iizuka:2021tut}
N.~Iizuka, A.~Miyata and T.~Ugajin, \emph{{A comment on a fine-grained
  description of evaporating black holes with baby universes}},
  \href{https://arxiv.org/abs/2111.07107}{{\ttfamily 2111.07107}}.

\bibitem{Nomura:2018kia}
Y.~Nomura, \emph{{Reanalyzing an Evaporating Black Hole}},
  \href{https://doi.org/10.1103/PhysRevD.99.086004}{\emph{Phys. Rev. D}
  {\bfseries 99} (2019) 086004}
  [\href{https://arxiv.org/abs/1810.09453}{{\ttfamily 1810.09453}}].

\bibitem{Maldacena:2016upp}
J.~Maldacena, D.~Stanford and Z.~Yang, \emph{{Conformal symmetry and its
  breaking in two dimensional Nearly Anti-de-Sitter space}},
  \href{https://doi.org/10.1093/ptep/ptw124}{\emph{PTEP} {\bfseries 2016}
  (2016) 12C104} [\href{https://arxiv.org/abs/1606.01857}{{\ttfamily
  1606.01857}}].

\bibitem{Harlow:2018tqv}
D.~Harlow and D.~Jafferis, \emph{{The Factorization Problem in
  Jackiw-Teitelboim Gravity}},
  \href{https://doi.org/10.1007/JHEP02(2020)177}{\emph{JHEP} {\bfseries 02}
  (2020) 177} [\href{https://arxiv.org/abs/1804.01081}{{\ttfamily
  1804.01081}}].

\bibitem{Cadoni:1995dd}
M.~Cadoni, \emph{{Trace anomaly and Hawking effect in generic 2-D dilaton
  gravity theories}},
  \href{https://doi.org/10.1103/PhysRevD.53.4413}{\emph{Phys. Rev. D}
  {\bfseries 53} (1996) 4413}
  [\href{https://arxiv.org/abs/gr-qc/9510012}{{\ttfamily gr-qc/9510012}}].

\bibitem{Mann:1992yv}
R.~B. Mann, \emph{{Conservation laws and 2-D black holes in dilaton gravity}},
  \href{https://doi.org/10.1103/PhysRevD.47.4438}{\emph{Phys. Rev. D}
  {\bfseries 47} (1993) 4438}
  [\href{https://arxiv.org/abs/hep-th/9206044}{{\ttfamily hep-th/9206044}}].

\end{thebibliography}\endgroup

\end{document}